\documentclass[usenatbib]{mn2e}
\usepackage{amsmath}
\usepackage{graphicx}
\usepackage{pgfplots}
\pgfplotsset{compat=1.5}

\def\lesssim{\mathrel{\hbox{\rlap{\hbox{\lower4pt\hbox{$\sim$}}}\hbox{$<$}}}}
\def\gtrsim{\mathrel{\hbox{\rlap{\hbox{\lower4pt\hbox{$\sim$}}}\hbox{$>$}}}}

\def\arcmin{\hbox{$^\prime$}}

\newcommand{\mamo}[1]{\mbox{$#1$}}
\newcommand{\unit}[1]{\ifmmode \:\mbox{\rm #1}\else \mbox{#1}\fi}
\newcommand{\sbr}[1]{_{\rm #1}}
\newcommand{\spr}[1]{^{\rm #1}}
\newcommand{\expec}[1]{\mamo{\left\langle #1 \right\rangle}}
\newcommand{\stderr}[1]{\mamo{\sigma_{ #1 }}}
\newcommand\numberthis{\addtocounter{equation}{1}\tag{\theequation}}
\newcommand{\km}{\unit{km}}
\newcommand{\s}{\unit{s}}

\newcommand{\kpc}{\unit{kpc}}
\newcommand{\mpc}{\unit{Mpc}}

\newcommand{\chisq}{\mamo{\chi^2}}
\newcommand{\msun}{\mamo{M_{\odot}}}

\newcommand{\Seclabel}[1]{\label{sec:#1}}

\newcommand{\Applabel}[1]{\label{sec:#1}}
\newcommand{\Eqlabel}[1]{\label{eq:#1}}
\newcommand{\Figlabel}[1]{\label{fig:#1}}

\newcommand{\Secref}[1]{Section~\ref{sec:#1}}

\newcommand{\Appref}[1]{Appendix~\ref{sec:#1}}
\newcommand{\Eqref}[1]{Equation~(\ref{eq:#1})}
\newcommand{\Figref}[1]{Fig.~\ref{fig:#1}}

\title[Measuring Weak Lensing Magnification With Weighted Number Counts]{A Generalized Method for Measuring Weak Lensing Magnification With Weighted Number Counts}
\author[B.R. Gillis and A.N. Taylor]{Bryan R. Gillis$^{1}$\thanks{E-mail: brg@roe.ac.uk}, Andy Taylor$^{1}$ \\
$^{1}$Institute for Astronomy, University of Edinburgh, Royal Observatory Edinburgh, Edinburgh, EH9 3HJ, United Kingdom.}

\begin{document}

\date{29 Jun 2015}

\pagerange{\pageref{firstpage}--\pageref{lastpage}} \pubyear{2015}

\maketitle

\label{firstpage}

\begin{abstract}

\noindent We present a derivation of a generalized optimally-weighted estimator for the weak lensing magnification signal, including a calculation of errors. With this estimator, we present a local method for optimally estimating the local effects of magnification from weak gravitational lensing, using a comparison of number counts in an arbitrary region of space to the expected unmagnified number counts. We show that when equivalent lens and source samples are used, this estimator is simply related to the optimally-weighted correlation function estimator used in past work and vice-versa, but this method has the benefits that it can calculate errors with significantly less computational time, that it can handle overlapping lens and source samples, and that it can easily be extended to mass-mapping. We present a proof-of-principle test of this method on data from the CFHTLenS, showing that its calculated magnification signals agree with predictions from model fits to shear data. Finally, we investigate how magnification data can be used to supplement shear data in determining the best-fit model mass profiles for galaxy dark matter haloes. We find that at redshifts greater than $z \sim 0.6$, the inclusion of magnification can often significantly improve the constraints on the components of the mass profile which relate to galaxies' local environments relative to shear alone, and in high-redshift, low- and medium-mass bins, it can have a higher signal-to-noise than the shear signal.

\end{abstract}


\section{Introduction}

Weak gravitational lensing has proven to be a powerful probe of the distribution of dark matter in the Universe, allowing the dark mass within clusters to be mapped, providing useful measurements of the halo masses of galaxies at a variety of redshifts, and helping to constrain fundamental cosmological parameters, among other applications. However, these advances have relied only on shear, which is just one of the two effects weak gravitational lensing has on background galaxies; the other being magnification.

When a background object is affected weakly by lensing, its shape will be stretched tangentially to the direction to the lens, known as shear, and its size will change, known as magnification. For measuring lensing, shear presents two key advantages over magnification: It has an expectation value of zero in the absence of lensing, and it typically has a larger signal-to-noise ratio than magnification. The zero unlensed expectation value of shear has many benefits which make measuring it easier. It allows shear to be measured without calibration to determine the unlensed expectation value, and it also indirectly leads to the result that most biases in measuring shear are multiplicative rather than additive. Any measurement of magnification, on the other hand, must deal with a comparison to a quantity that has a non-zero expectation value, which allows additive biases to play a role. As such, magnification requires much more care in analysis in order to determine an unbiased estimator.

A relatively simple calculation of signal-to-noise which assumes comparable shear and convergence \citep[as done by eg.][]{SchKinErb00} can show that shear is generally expected to provide much stronger signal-to-noise than magnification at low redshift. However, the benefit from magnification is not entirely negligible, and no extra observing time is needed to gain the information required for a magnification measurement; any observation which can measure shear can also measure magnification. As such, if an unbiased measurement of it can be made, it will at least provide a slight improvement to weak lensing mass measurements. The benefits of magnification are more significant at higher redshifts, due to the fact that the number of galaxies sufficiently resolved such that their shapes can be measured drops off faster with redshift than the total number of observed galaxies. For lens galaxies at redshifts much greater than $\sim 1$, the shear signal is negligible, but the magnification signal is still sufficient to gain some information on the masses of these galaxies \citep{vanHilFor10,HilvanSco13}. Magnification can also provide additional information if the convergence and shear in a region of space are not of the same magnitude. For instance, as shear is only sensitive to changes in overdensity, a region of constant projected mass density could have positive convergence but no shear, allowing magnification to be used to break the ``mass-sheet degeneracy'' that prevents shear alone from determining this overdensity.

It is thus well-motivated to further investigate the uses of weak lensing magnification in measuring the mass distributions of dark matter haloes. This has been shown to be possible in recent works \citep[eg.][]{ScrMRic05,HilPieErb09,MScrFuk10,ForHilVan12,HilvanSco13,ForHilVan14}, although the results so far are limited, particularly compared to the wealth of results which have been obtained from shear data. These recent measurements of magnification have frequently used an optimally-weighted correlation function estimator \citep{MBar02,ScrMRic05}, which compares the projected number density of source galaxies in the vicinity of lens galaxies to a random catalogue, properly weighted by a function of the luminosities of the source galaxies to optimize the magnification signal.\footnote{There is also notable work using a conceptually-different approach to magnification, relying on the observed properties of galaxies to predict their intrinsic luminosities \citep[eg.][]{HufGra14}. It remains for future work to investigate if it is possible to combine the results of this approach with the type of approach presented in this paper.} The measured correlation function is then compared to the correlation function predicted from models. While this methodology is sufficient for purposes such as galaxy-galaxy lensing, it cannot easily be applied to other lensing applications such as mass mapping. It also has the drawbacks that it requires significant computational time to calculate errors through a jackknife or bootstrap approach, and it requires a clear separation between lens and source samples to avoid contamination from clustering.

In this paper, we start from the assumption that we desire an optimally-weighted estimator for the magnification itself, as opposed to previous work which has used an optimized correlation function estimator, and we present a derivation of such an estimator. Using this estimator, we present a new method for measuring the magnification signal which avoids some of the drawbacks of the optimally-weighted correlation function, and we show how their resulting estimators are related. We present a proof-of-principle of this new method, and we look at the results of this test to analyse the possible benefits magnification data may provide in supplementing shear data.

In \Secref{Methodology} of this paper, we present a derivation of an estimator for magnification and a derivation of errors for it, and we discuss how it can be implemented in practice for galaxy-galaxy lensing. In \Secref{Data}, we discuss the data we use for testing our method, which we take from the CFHTLenS. In \Secref{Testing}, we demonstrate that our method can indeed detect magnification signals, and we look at the benefits of supplementing shear data with magnification data for galaxy-galaxy lensing. In \Secref{Discussion}, we discuss various aspects of the work in this paper, including a comparison of our method and the optimally-weighted correlation function method, and we present a proposed implementation of our method for mass-mapping. In \Secref{Conclusions} we summarize and present our conclusions.

In order to facilitate future work with magnification, we make the code used for this paper publicly-available. Details of its access can be found in \Appref{code_access}.

For relevant calculations in this paper, we assume a cosmology with ${\rm H}\sbr{0} = 70 \km\s^{-1}\mpc^{-1}$, $\Omega\sbr{m} = 0.288$, $\Omega\sbr{r} = 8.6\times 10^{-5}$, and $\Omega\sbr{k}=0$. Unless stated otherwise, all mass values are in units of solar masses and all distance values are in units of $\kpc$.

\section{Magnification Measurement Methodology}
\Seclabel{Methodology}

In this section we present a derivation of a generalized optimally-weighted estimator for magnification and our proposed method for measuring it. In \Secref{estimator_math}, we present the derivation of this estimator and show how its errors can be determined. In \Secref{galgal_implementation}, we present our method for measuring magnification for the case of galaxy-galaxy lensing. Finally, in \Secref{determining_source_density}, we discuss the key step of determining the background, unmagnified source density.

\subsection{Determining an Optimal Local Estimator for Magnification}
\Seclabel{estimator_math}

Let us start with the assumption that we wish to measure the mean magnification $\mu$ in a patch of the sky where $\mu$ is expected to be approximately constant, such as in a circular annulus around a lens galaxy. We first bin the galaxies observed in this patch at redshifts greater than that of the lens by magnitude, using small bins of width $dm$, which we will later decrease to the limit of zero. Let
\begin{equation}
dn_i = n(m_i)dm
\end{equation}
be the number of galaxies observed in the $i\spr{th}$ magnitude bin in this patch of sky, and let
\begin{equation}
dn_{{\rm 0},i} = n_{\rm 0}(m_i)dm
\end{equation}
be the expected number of galaxies in this magnitude bin, in a patch of sky of the same area, and in the same redshift range in the presence of no magnification. If we assume that the number count can be locally approximated by a power law with slope
\begin{equation}
\alpha_i = 2.5 \frac{d}{dm_i}\log n_{\rm 0}(m_i)\mathrm{,}
\Eqlabel{alpha_definition}
\end{equation}
then we expect the observed number count of galaxies in this magnitude bin to be:
\begin{equation}
\left<dn_i\right> = dn_{{\rm 0},i}\mu^{\alpha_i-1}
\Eqlabel{mu_definition}
\end{equation}
\citep{Nar89,BroTayPea95}.

In the weak lensing limit, we expect $(\mu-1) \ll 1$, and so we can use the approximation:
\begin{align*}
\expec{dn_i} &= dn_{{\rm 0},i}\left(1+(\mu-1)\right)^{\alpha_i-1} \\ \numberthis
&\approx dn_{{\rm 0},i}\left(1+(\alpha_i-1)(\mu-1)\right)\mathrm{.}
\end{align*}
Solving this equation for $\mu$, we get:
\begin{equation}
\mu = 1+\frac{\expec{dn_i}-dn_{{\rm 0},i}}{(\alpha_i-1)dn_{{\rm 0},i}}\mathrm{.}
\end{equation}
If we assume the statistical error on estimating $dn_{{\rm 0},i}$ and $\alpha_i$ to be significantly less than the standard error of $dn_i$, we can then use
\begin{equation}
\mu_i = 1+\frac{dn_i-dn_{{\rm 0},i}}{(\alpha_i-1)dn_{{\rm 0},i}}
\Eqlabel{mu_i_final}
\end{equation}
as an estimator for $\mu$, with $\expec{\mu_i} = \mu$, and which is independent for each magnitude bin $i$.

In order to optimally combine the estimators $\mu_i$ for all magnitude bins, we apply inverse-variance weighting. We start with the standard deviation of each estimator, which, under the assumption that $dn_i$ is Poisson, can be calculated to be:
\begin{equation}
\Eqlabel{mu_stddev}
\sigma_{\mu_i} \approx \frac{\sqrt{1+(\alpha_i-1)(\mu-1)}}{\left|\alpha_i-1\right|\sqrt{dn_{{\rm 0},i}}}\mathrm{.}
\end{equation}
The derivation of this equation is presented in \Eqref{mu_stddev_derivation} in \Appref{derivations}. This then gives us the optimal inverse-variance weighting for bin $i$:
\begin{equation}
\Eqlabel{opt_weighting}
dw_{i,{\rm opt}} = \frac{1}{\sigma_{\mu_i}^2} = \frac{dn_{{\rm 0},i}\left(\alpha_i-1\right)^2}{1+\left(\alpha_i-1\right)\left(\mu-1\right)}\mathrm{.}
\end{equation}
While the weighting scheme in \Eqref{opt_weighting} will provide the estimate of $\mu$ with the minimum standard error, it requires foreknowledge of the actual value of $\mu$ to determine the weights. For a likelihood-based approach, we can use the value of $\mu$ predicted by the model for the weight calculation, but this will require a separate summation over bins to calculate the likelihood for each model $\mu$. Since we expect $(\mu-1) \ll 1$, we instead use the weighting scheme:
\begin{equation}
dw_{i} = dn_{{\rm 0},i}\left(\alpha_i-1\right)^2\mathrm{,}
\Eqlabel{weighting_scheme}
\end{equation}
which is model-independent. This will save significant computational time at the expense of a negligible decrease in the precision (but not accuracy) of our constraints.

We can then calculate our estimate of $\mu$ for this patch of sky through:
\begin{equation}
\hat{\mu} = \frac{\sum_{i} dw_i \mu_i}{\sum_{i} dw_i} \mathrm{.}
\end{equation}
This will have standard error:
\begin{equation}
\stderr{\hat{\mu}} \approx \left(\sum_{i} dw_i\right)^{-1/2} \mathrm{,}
\end{equation}
the accuracy of which depends on the accuracy of the standard deviation we calculated for each $\mu_i$ in \Eqref{mu_stddev}.

Let us now examine the case where the bin size $dm$ approaches zero. We now express the parameters $dn_i$, $dn_{{\rm 0},i}$, $\alpha_i$, $\mu_i$, and $dw_{i}$ as $dn(m)$, $dn_{\rm 0}(m)$, $\alpha(m)$, $\mu(m)$, and $dw(m)$ respectively. Here, $m$ is the magnitude of the differential bin, and we define $dn(m) = n(m)dm$, $dn_{\rm 0}(m) = n_{0}(m)dm$, and $dw(m)=w(m)dm$. $n_{0}(m)$ and $\alpha(m)$ have functional forms which can be constrained by observations, also giving us a functional form for $w(m)$. For a set of observed galaxies with magnitudes $m\sbr{1}, m\sbr{2}... m_j$, we can express the differential number count $n(m)$ in terms of a sum of Dirac delta functions:
\begin{equation}
  n(m) = \sum_j \delta(m-m_j)\mathrm{.}
  \Eqlabel{n_of_m_differential_form}
\end{equation}
This form gives the expected behaviour that integrating $n(m)$ from $m$ to $m+dm$ results in the number of galaxies with magnitudes in this range.

Let $a$ represent the minimum observable magnitude for galaxies, and let $b$ represent the maximum magnitude used in the catalogue. We can then calculate our estimate of $\mu$ through:
\begin{align*}
  \hat{\mu} &= \frac{1}{W} \int_{a}^{b} w(m)\mu(m)dm \\   \numberthis
 &= 1 + \frac{1}{W} \int_{a}^{b} \left[n(m)-n_{0}(m)\right]\left[\alpha(m)-1\right]dm
 \Eqlabel{mu_hat_integral_form}
\end{align*}
where:
\begin{equation}
  W = \int_{a}^{b}w(m)dm = \int_{a}^{b}n\sbr{0}(m)\left(\alpha(m)-1\right)^2 dm\mathrm{.}
 \Eqlabel{mu_W_definition}
\end{equation}
Expanding $n(m)$ using \Eqref{n_of_m_differential_form}, this gives us:
\begin{align*}
 \Eqlabel{mu_hat_sum_form}
  \hat{\mu} &= 1 + \\ \numberthis
  &\frac{1}{W} \left( \sum_j\left[\alpha(m_j)-1\right] - \int_{a}^{b} n_{0}(m)\left[\alpha(m)-1\right]dm \right)\mathrm{,}
\end{align*}
where $m_j$ represent the magnitudes of observed galaxies in the range $a < m_j < b$.

As before, this estimator will have the standard error:
\begin{equation}
  \stderr{\hat{\mu}} \approx W^{-1/2}\mathrm{,}
 \Eqlabel{mu_hat_stderr}
\end{equation}
and will be approximately Gaussian for large $n$. The accuracy of this depends on the validity of our assumptions on the distribution of number counts used in \Eqref{mu_stddev}. In practice, an empirical error estimate (determined, for instance, from the standard deviation in the $\hat{\mu}$ for various similar regions) is likely to be more reliable.

From this, we can estimate the likelihood for a given value of $\mu$ to be
\begin{align*}
 \Eqlabel{mu_hat_likelihood}
  \mathcal{L}(\mu|\hat{\mu}) &= \frac{1}{\sqrt{2\pi}\sigma_{\hat{\mu}}}\exp\left( -\frac{\mu-\hat{\mu}}{2\sigma_{\hat{\mu}}^2} \right) \\ \numberthis
  &\approx \sqrt{\frac{W}{2\pi}}\exp\left( -\frac{W (\mu-\hat{\mu})}{2} \right) \mathrm{.} \\ \nonumber
\end{align*}

The weighting used for this estimator is in fact the same as that generally used for the optimally-weighted correlation function, and was originally presented by \citet{MBar02} without derivation. We will later show in \Secref{method_comparison} that this estimator is simply related to that used for correlation-function-based methods in the scenario in which the lens and source samples are equivalent. In the following section, we will show how the estimator presented here can be used as a basis for an alternative method for calculating a magnification signal by directly comparing the observed source counts to the measured background density at a given redshift.

\subsection{Implementation for Galaxy-Galaxy Lensing}
\Seclabel{galgal_implementation}

We start with the case of a galaxy-galaxy lensing analysis, which will allow us to readily test our implementation against the lensing signal measured with shear information. We start by binning our lens sample by both stellar mass and redshift so that the lenses in each bin will have approximately the same halo mass profiles.

For each lens, we bin the possible patch of sky in which sources could potentially appear based on projected distance at the redshift of the lens. This gives us a series of annuli for each lens, some of which may overlap masked regions of the image. We determine the unmasked fraction of each annulus through the ratio of unmasked pixel centres to all pixel centres inside the annulus. In practice, most masks are relatively coarse, and so this fraction can be reasonably estimated by regularly sampling pixels in order to conserve computational time.

The angular area of the $k\spr{th}$ annulus around the $l\spr{th}$ lens is then:
\begin{equation}
	\Omega_{kl} = D_{l}^{-2}f_{kl}\pi\left( R_{k+1}^{2} - R_{k}^{2}\right)\mathrm{,}
\end{equation}
where $D_{l}$ is the angular diameter distance at the redshift of the lens, $f_{kl}$ is the unmasked fraction of this annulus, and $R_{k}$ and $R_{k+1}$ are its inner and outer radii respectively in physical units.

To ensure minimal contamination of the source sample with galaxies which are actually in the lens plane, we require any sources used for our analyses to be separated from their associated lenses by a redshift buffer $\Delta z$, which we set to $0.2$. As galaxies in the lens plane are clustered with each other, any which are misidentified as sources will be clustered with lenses as well, resulting in an artificial increase in the observed counts of sources around lenses, biasing the magnification signal. It is therefore necessary that this buffer be sufficiently large that this contamination has negligible effect on the measured magnification signal. See \Secref{galgal_sample_selection} for discussion of our choice of $\Delta z$.

Note that this separation between lenses and sources is applied to each lens individually. This allows for the possibility that the lens sample might overlap with the source sample. Previous methods for measuring magnification have relied on the two samples being well-separated. The loosening of this requirement for our method will allow for an increased signal-to-noise ratio in situations in which it is beneficial to use overlapping lens and source samples.

The expected unmagnified number count of sufficiently-distant sources in a given magnitude bin projected in this annulus is
\begin{equation}
	n_{{\rm 0},kl}(m)dm = \Omega_{kl} n(m,z+\Delta z)dm\mathrm{,}
\end{equation}
where $z$ is the redshift of the lens and $n_{\rm 0}(m,z+\Delta z)dm$ is the number of sources per unit area with magnitude between $m$ and $m+dm$ at redshift $z+\Delta z$ or greater. This gives us the relevant $n\sbr{0}(m)$ to use in \Eqref{mu_hat_sum_form}, and $\alpha(m)$ can be determined through \Eqref{alpha_definition}, allowing us to get the estimate $\hat{\mu}_{kl}$ for this annulus.

This process can be performed more efficiently through storing the integrals
\begin{equation}
	d(z) = \int_{a}^{b} n(m,z)\left[\alpha(m)-1\right] dm
\end{equation}
and
\begin{equation}
	w(z) = \int_{a}^{b} n(m,z)\left[\alpha(m)-1\right]^2 dm
\end{equation}
for all needed redshift values and calculating $\hat{\mu}_{kl}$ through:
\begin{align*}
	\Eqlabel{mu_hat_efficient_form}
	&\hat{\mu}_{kl} = \\ \numberthis
	&1 + \frac{1}{\Omega_{kl} w(z\sbr{l}+\Delta z)}\left( \sum_{j}\left[\alpha(m_{jkl})-1\right] - \Omega_{kl} d(z\sbr{l}+\Delta z) \right) \mathrm{,}
\end{align*}
where $m_{jkl}$ represent the magnitudes of all sufficiently-distant sources observed in this annulus, and $a$ and $b$ represent the lower and upper magnitude limits of the source sample, respectively.

By comparing \Eqref{mu_hat_efficient_form} to \Eqref{mu_hat_integral_form} and \Eqref{mu_hat_stderr}, we can estimate the standard deviation of $\hat{\mu}_{kl}$ by
\begin{equation}
	\sigma_{\hat{\mu}_{kl}} = \left[ \Omega_{kl} w(z\sbr{l}+\Delta z) \right]^{-1/2}\mathrm{.}
	\Eqlabel{sigma_mu_hat_efficient_form}
\end{equation}
When we stack lenses, we can use this estimate to apply inverse-variance weighting, giving each individual annulus a weight of
\begin{equation}
	w_{kl} = \Omega_{kl} w(z\sbr{l}+\Delta z)\mathrm{.}
	\Eqlabel{weight_mu_hat_efficient_form}
\end{equation}

Finally, we get the estimated magnification for the stacked lenses in this annulus from the weighted mean across all lenses:
\begin{equation}
	\hat{\mu}_k = \frac{ \sum_l w_{kl} \hat{\mu}_{kl} } { \sum_l w_{kl}  }\mathrm{.}
	\Eqlabel{mu_hat_stacked}
\end{equation}
If our estimates of standard deviations in number counts were perfectly accurate, the standard error in this estimator would be
\begin{equation}
	\stderr{\hat{\mu}_k,{\rm ideal}} = \left( \sum_l w_{kl} \right)^{-1/2}\mathrm{,}
	\Eqlabel{mu_hat_stacked_error}
\end{equation}
but as this is unlikely to be the case (for instance, due to the clustering of sources making their distribution not perfectly Poisson), we instead determine it empirically. We start by empirically determining the weighted standard deviation in the $\hat{\mu}_{kl}$ estimates for the set of lenses,
\begin{equation}
	\sigma_{\hat{\mu}_{k}} = C_k\sqrt{ \frac{ \sum_l w_{kl} \hat{\mu}_{kl}^2 } { \sum_l w_{kl} } - \left( \frac{ \sum_l w_{kl} \hat{\mu}_{kl} } { \sum_l w_{kl} } \right)^2 }\mathrm{,}
	\Eqlabel{mu_stderr_1}
\end{equation}
where
\begin{equation}
	C_k = \frac{\sum_l w_{kl}}{\sqrt{\left(\sum_l w_{kl}\right)^2 - \sum_l w_{kl}^2}}
	\Eqlabel{mu_stderr_2}
\end{equation}
\citep{ThePawn15}\footnote{This formula for the standard error of a weighted mean was chosen based on a comparison of various estimators, as it was found to generally have the lowest amount of bias in various test cases, some of which involved biased weights being used.}. We then estimate the standard error in the magnification estimate $\hat{\mu}_k$ for this set of annuli through:
\begin{equation}
	\stderr{\hat{\mu},k} = \frac{\sigma_{\hat{\mu}_{k}}}{ \sqrt{ N_{k}-1 }}\mathrm{,}
	\Eqlabel{mu_stderr_3}
\end{equation}
where $N_k$ is the number of lenses with non-zero weight for this annulus.

This estimate of the magnification can then be converted into an estimate of the convergence:
\begin{equation}
	\hat{\kappa}_k = 1 - \sqrt{ \hat{\mu}_k^{-1} + \hat{\gamma}_{{\rm t},k}^2 },
	\Eqlabel{kappa_estimate}
\end{equation}
where $\hat{\gamma}_{{\rm t},k}$ is the tangential-shear estimate for this annulus\footnote{\Eqref{kappa_estimate} uses the assumption that the cross-shear $\gamma_{{\rm x},k}$ is expected to be zero, as is the case in galaxy-galaxy lensing. For applications where this is not the case, $\gamma_k = \sqrt{ \gamma_{{\rm t},k}^2 + \gamma_{{\rm x},k}^2 }$ should be used in place of $\gamma_{{\rm t},k}$ in this equation and \Eqref{kappa_stderr}. }. The standard error in this estimate is
\begin{equation}
	\stderr{\hat{\kappa},k} = \sqrt{\frac{\stderr{\hat{\mu}}^2 + 2 \hat{\gamma}_{{\rm t},k} \stderr{\hat{\gamma}_{{\rm t},k}^2}}{4\left(\hat{\mu}_k^{-1} + \hat{\gamma}_{{\rm t},k}^2\right)}}\mathrm{.}
	\Eqlabel{kappa_stderr}
\end{equation}
The estimated projected overdensity in this annulus is then
\begin{equation}
	\hat{\Sigma}_k = \Sigma_{{\rm crit},k}\hat{\kappa}_k\mathrm{,}
\end{equation}
where $\Sigma_{{\rm crit},k}$ is the critical density for this annulus,
\begin{equation}
	\Sigma_{{\rm crit},k} = \frac{c^2}{4\pi G\sbr{c}}\frac{D\sbr{s}}{D\sbr{l}D\sbr{ls}}\mathrm{,}
	\Eqlabel{Sigma_crit_def}
\end{equation}
where $D\sbr{s}$ is the angular diameter distance from redshift zero to the mean source redshift, $D\sbr{l}$ from redshift zero to the mean lens redshift, and $D\sbr{ls}$ from the mean lens redshift to the mean source redshift. The standard error in the projected overdensity is then:
\begin{equation}
\stderr{\hat{\Sigma},k} = \Sigma_{{\rm crit},k}\stderr{\hat{\kappa},k}\mathrm{.}
\end{equation}

\subsection{Determining Unmagnified Source Density}
\Seclabel{determining_source_density}

In measuring the magnification signal of a given region of space, we require knowledge of what the source density would be as a function of magnitude in the absence of any magnification. In \Eqref{mu_hat_integral_form} and \Eqref{mu_hat_sum_form}, this represents $n\sbr{0}(m)$, which is used to determine $\alpha(m)$ through \Eqref{alpha_definition} and $W$ through \Eqref{mu_W_definition}. Given the large area covered by the CFHTLenS, it is a relatively safe assumption that the survey region is on average unmagnified. There will be some overdense and underdense regions, but the region as a whole should have a projected density approximating that of the whole sky, and so we can thus use the measured source densities across the entire field to determine $n\sbr{0}(m)$.

This approach has two key advantages over a theoretical model. First, it accounts for many observational effects which will identically affect the source samples we later use for measuring magnification, such as the difficulty in identifying faint galaxies. Second, as we are interested in all galaxies which appear behind a given annulus, we are in fact interested in the number density of all galaxies at a given redshift or greater, whereas models are typically fit only to galaxies within a single redshift slice. This resulted in us being unable to satisfactorily fit a model such as a Schechter function to the observed galaxy counts beyond a given redshift, even allowing for perturbations to the functional form.

This approach does result in noise being present in $n\sbr{0}(m)$. As it is measured over a much greater area of the sky than even the stacked regions we wish to measure magnification for, however, this noise is likely to be sub-dominant to the Poisson noise in $n(m)$, the number count distribution of galaxies which actually appear in these regions. However, we must differentiate $n\sbr{0}(m)$ to determine $\alpha(m)$, and differentiation amplifies this noise. In order to handle this, we first smooth the measured $n\sbr{0}(m)$ using the \citet{SavGol64} algorithm prior to differentiation.

As can be seen most easily in \Eqref{mu_hat_integral_form}, the estimator for magnification is particularly sensitive to $n\sbr{0}$. For instance, in the simplified scenario where $\alpha(m) = 2$, a bias of $0.1$ per cent in $n\sbr{0}$ leads directly to a bias of of $-0.001$ in $\hat{\mu}$. As the typical magnification signal is of this order, we must take care to minimize bias in $n\sbr{0}$ as much as possible.

If the positions of lenses and sources were completely independent, the approach we use would be expected to be unbiased when applied to galaxy-galaxy lensing. However, even aside from contamination of the source sample with galaxies in the lens plane misidentified as sources, some correlation or anti-correlation between the samples is expected. This can be caused by a number of position-dependent effects which can introduce biases into the redshift estimates for galaxies, including atmospheric seeing, depth variations due to the dithering strategy of the survey, and depth variations due to galactic dust. 

The method used for estimating photometric redshifts in the CFHTLenS by \citet{HilErbKui12} used a local model for atmospheric seeing to minimize the bias in galaxy colours due to it, so this is unlikely to be an issue. However, regions with poor seeing and regions with poor depth will have larger errors on the estimated colours of galaxies. If there is a gradient in the object density in colour space (eg. more blue galaxies than red), then noise in colour determinations will result in asymmetric scatter along this gradient. Even if any particular galaxy is as likely to be scattered to a higher colour as a lower colour, the net result will be to smooth out the object density distribution in each axis of colour space. This will then result in the estimated redshifts of these objects being systematically biased. As the relation between redshift and colour is not linear, the direction and magnitude of this bias will vary with redshift. This means that some redshift bins may be spuriously underdense, while others are spuriously overdense.

Let us present an example of what might occur, and how this might lead to a systematic bias in the magnification signal. Consider an extreme example: a field is divided into two regions of equal size; one region has good seeing, and another region has poor seeing. This poor seeing results in the measured luminosities of galaxies in this region having greater errors than the luminosities of galaxies in the other region. If we suppose that there are more blue galaxies than red and that, for the sake of this example, the distribution between the two extremes is smooth, then we would expect more galaxies to be scattered to redder colours than we would expect to see galaxies scattered to bluer colours. Thus, there would be a net bias in this region for all galaxies to appear redder. Let us assume a simple photometric redshift estimation which relies only on this colour. In this case, this would result in a systematic bias in this region for all galaxies to have higher redshifts. As there is more volume at higher redshifts, there are more galaxies there. Thus, a bias of all galaxies to higher redshifts would result in a lower density of galaxies in any given redshift bin.

We thus have one region with good seeing and approximately normal density of lenses and sources, and one with poor seeing and a lower density of both lenses and sources. In measuring $n\sbr{0}$, we would weight both of these regions equally and calculate an expected source density which is the mean of the two. However, in measuring the magnification signal, we would be using the lower number of lenses in the region which also has a lower number of sources. Our measured $n$ would thus be more akin to a weighted mean, which in this case would end up being weighted more by the region with a lower density of sources. Thus, even without any magnification, we have a scenario in which the measured count of galaxies differs from the expected count.

\begin{figure}
	\includegraphics[scale=0.45]{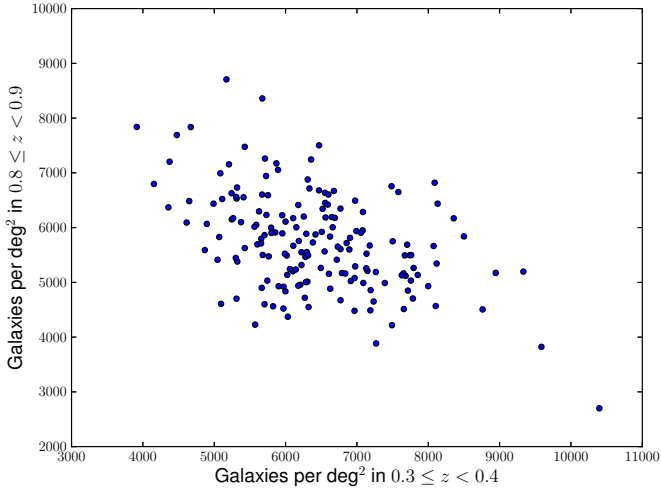}
	\caption[Comparison of densities of galaxies in two different redshift slices.]{A comparison of the number densities of galaxies in two different photometric redshift slices in the CFHTLenS. Each point represents an individual field. The anti-correlation between the two redshift slices implies the presence of a systematic error in redshift estimates which varies from field to field.}
	\Figlabel{CFHTLenS_field_densities}
\end{figure}

Evidence of this type of bias can be seen in \Figref{CFHTLenS_field_densities}, which shows the galaxy number densities of different fields from the CFHTLenS in the redshift slices $0.3-0.4$ and $0.8-0.9$ plotted against each other. This plot shows clearly that fields which have greater densities in the $0.3-0.4$ slice have lower densities in the $0.8-0.9$ slice, and vice-versa. There are also many case of well-separated slices which have correlated number densities, but the anti-correlation here is most striking, as it is difficult to explain with any effect other than non-trivial redshift systematic errors such as we have discussed. More insight into this can be gained from the rightmost panel of Figure 4 of \citet{HilErbKui12}, which shows a comparison of the photometric redshifts as used here against the spectroscopic redshifts of a matched sample of galaxies, and observing the complicated shape of the distribution. The field-by-field comparison we show here demonstrates that this distribution is not constant between fields.

A full discussion of this type of effect is beyond the scope of this paper. Interested readers are directed to Morrison and Hildebrandt (2015, submitted), which discusses this issue in detail.

In the example above of how this effect might lead to an error in the measured magnification signal, we pointed out that the result is similar to the difference between the unweighted and weighted means of source densities. We can thus correct for it by weighting our initial calculation of mean source density by the density of lenses in the same region of space. The more local this weight is, the noisier it will be, and the more it will become entangled with the actual magnification signal, but the better it will account for this effect.

For this paper, we have decided to weight on a field-by-field basis, as the discontinuity in both position and time of observation between fields is likely to result in much more variation in seeing and depth than the variation across the field-of-view within a single field. To do this, when we are counting all galaxies in an individual field at a redshift $z\sbr{s}$ or greater, we weight this by the number of lenses at redshift\footnote{The photometric redshifts we use are binned to discrete intervals of $0.01$, hence the precise value being used here.} $z\sbr{s}-\Delta z$. See \Secref{Field_weighting_effects} for discussion of the impact of this weighting on our results.

\section{Data}
\Seclabel{Data}

In order to test our methodology, we use the publicly-available galaxy catalogues from the Canada-France-Hawaii Telescope Lensing Survey \citep{HeyvanMil12}, hereafter referred to as ``CFHTLenS,'' which are available for download from \texttt{http://www.cadc-ccda.hia-iha.nrc-cnrc.gc.ca/ community/CFHTLens/query.html}. The CFHTLenS is a 154 deg$^2$ survey (125 deg$^2$ after masking) \citep{ErbHilMil13}, based on the Wide component of the Canada-France-Hawaii Telescope Legacy Survey, which was observed in the period from March 22nd, 2003 to November 1st, 2008, using the MegaCam instrument \citep{BouChaAbb03}. It consists of deep, sub-arcsecond, optical data in the five optical SDSS-like filters $u^{*}g'r'i'z'$. The data is divided into 171 individual patches of sky, which we refer to as ``fields'' in this paper.

In order to avoid contamination from foreground stars and galaxies, cosmic rays, and other observational defects, masks were applied to the CFHTLenS fields using the
\texttt{automask} tool \citep{DieErbLam07,ErbHilLer09}. Shapes of background galaxies with $i' < 24.7$ in the unmasked regions were measured with the \textit{lens}fit shape measurement algorithm \citep{MilHeyKit12}, giving an effective density for sources with shapes of 11/arcmin$^2$ in the redshift range $0.2 < z\sbr{phot} < 1.3$ \citep{HeyvanMil12}. For the analysis in this paper, we use all fields in the survey, not simply those that passed the systematics tests for cosmic shear measurements \citep{HeyvanMil12}, as it has been demonstrated that fields with systematics that may affect cosmic shear have no effect on galaxy-galaxy lensing measurements \citep{VelvanHoe14}.

Photometric redshifts for the survey were estimated with the \textit{BPZ} code \citep{Ben00}. The accuracy of photo-z measurements for the CFHTLS-Wide survey was improved through homogenizing the PSF through different bands in the CFHTLenS survey. Photo-zs were made available for the entire survey, with a typical redshift uncertainty of  $\sim0.04(1+z)$ in the redshift range $0.2 < z\sbr{phot} < 1.3$ \citep{HilErbKui12}.

We use the stellar mass estimates described by \citet{VelvanHoe14}, obtained by fitting spectral energy distribution (SED) templates, following the method of \citet{IlbSalLe10}. These stellar masses were found to be in general agreement with deeper data such as WIRDS, which includes NIR filters \citep{BieHudMcC12}, up to $z = 0.8$.

\begin{figure*}
	\includegraphics[scale=0.45]{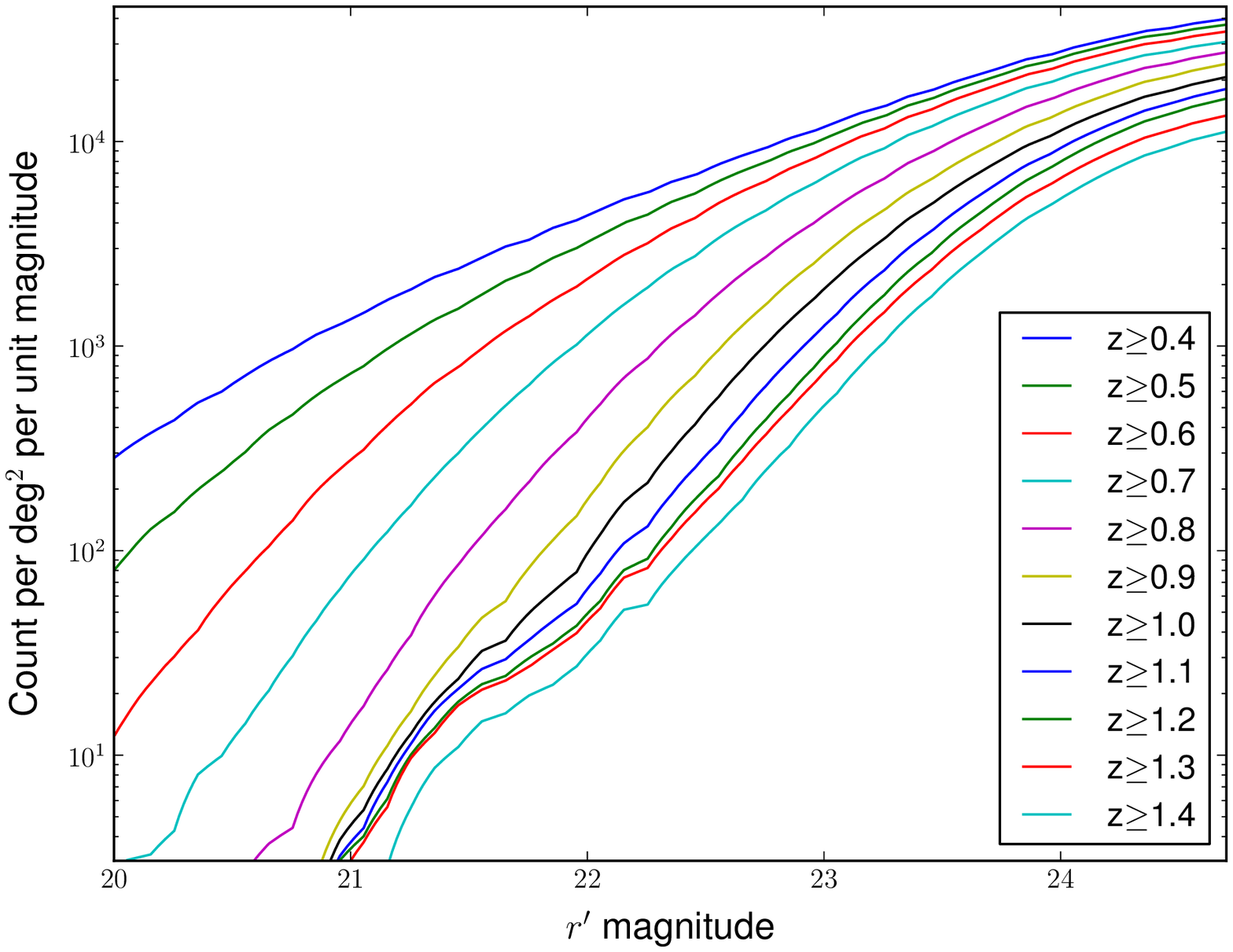}
	\includegraphics[scale=0.45]{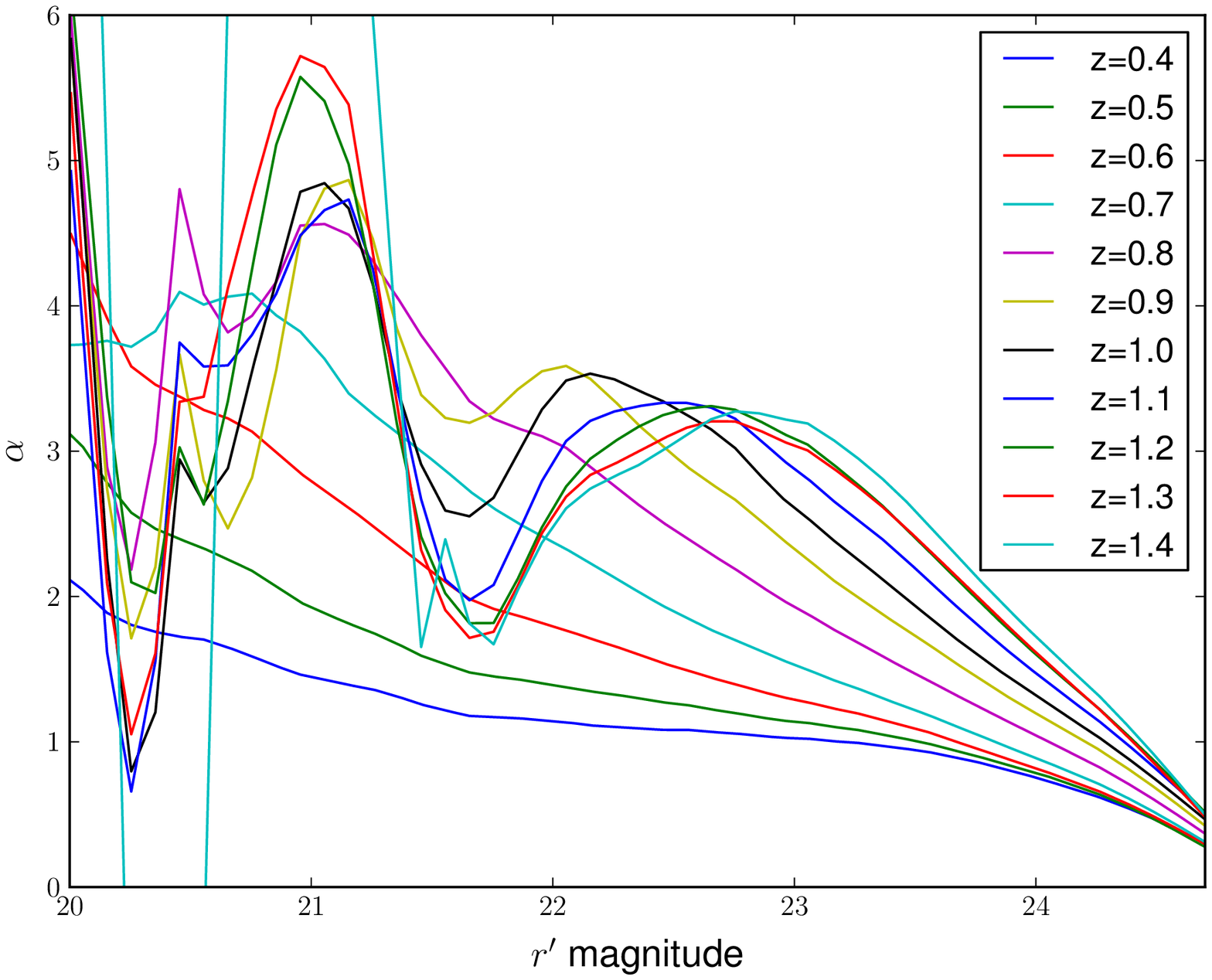}
	\caption[Number density of source galaxies vs magnitude.]{Number density of galaxies at a given redshift or greater as a function of $r'$-band magnitude (left) for the CFHTLenS, and the corresponding calculated values of $\alpha$ (right). The strong oscillations observed in $\alpha$ for high redshifts and low magnitudes are due to Poisson noise, due to the low numbers of galaxies in these bins. This results in these bins being correspondingly down-weighted, and so these oscillations are likely to have no appreciable effects on our results. In the left panel, the lowest redshifts correspond to the highest lines, and in the right panel, the lowest redshifts typically correspond to the lowest lines.}
	\Figlabel{ex_count_and_alpha}
\end{figure*}

We use the magnitudes measured with the $r'$ filter for all galaxies. We considered using the $i'$-band magnitudes, but this was found to be problematic, as the number density of galaxies as a function of $i'$-band magnitude in the CFHTLenS shows a prominent step at $m_{i'}=23$, which would result in $\alpha(m)$, the logarithmic derivative of this function, being discontinuous at this value. This is possibly an indirect result of the fact that the original $i'$ filter used for the CFHTLenS had to be replaced partway through observations. The $r'$-band magnitudes showed no such issue, and typically had the next best signal-to-noise after the $i'$-band magnitudes, and so we decided to use them. From here on, whenever we refer to magnitudes, we will thus be referring specifically to the $r'$-band magnitudes.

\subsection{Sample Selection}
\Seclabel{galgal_sample_selection}

We select galaxies from the public catalogue using the following cuts on it:
\begin{itemize}
	\item Z\_B $ \leq 4$ 
	\item MAG\_R $ \leq 24.7$
	\item star\_flag $ \leq 0.01$
	\item CHI\_SQUARED\_BPZ $ \leq 2$.
\end{itemize}
This provides us with a sample of galaxies with reasonable-quality redshifts. Other cuts were tested to improve redshift quality, such as imposing a cut on the ``ODDS'' parameter, but they were found to cause too much bias in the resulting galaxy selection. We use Z\_B as the photometric redshift estimate for each galaxy.

We do not impose a cut on the ``MASK'' parameter. Rather, we download the mask files and apply them ourselves, removing any galaxy from the sample which resides in a region with MASK $ > 1$. We do this as we found that the catalogue's MASK values for a small number of galaxies did not agree with our determinations, possibly due to a different version of the mask files having been used to assign the values in the catalogue. Assigning the MASK values ourselves thus ensures consistency with the mask files we have available, which we also use for determining the unmasked fractions of annuli around lenses.

\begin{figure}
	\includegraphics[scale=0.45]{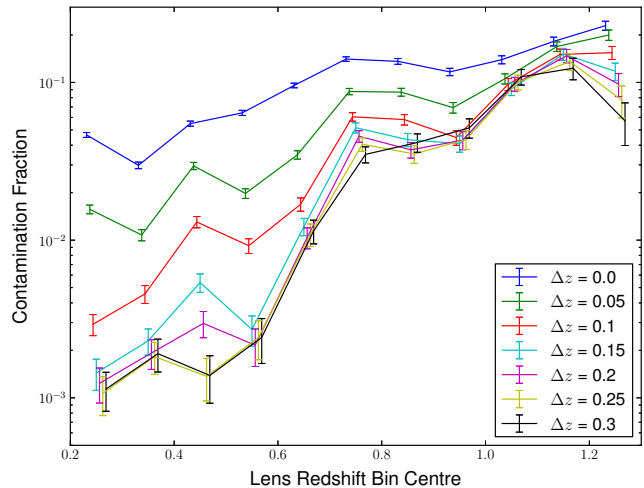}
	\caption[Source sample contamination fraction versus redshift.]{For a given lens redshift bin of width $0.1$, the fraction of galaxies with photometric redshifts at least a given redshift $\Delta z$ greater than the centre of the bin (corresponding to the source sample), which are likely to actually reside in the bin. We determine this fraction through matching galaxies in our sample with spectroscopic redshift catalogues from the VVDS and DEEP2, and determining the fraction of galaxies with sufficiently-large photometric redshifts which have spectroscopic redshifts within a given bin. Larger $\Delta z$ values consistently correspond to smaller contamination fractions, though this effect seems to converge to a maximum by $\Delta z \sim 0.2$. The contamination fraction values for each lens redshift bin are plotted near the centres of the bins, with a horizontal offset applied for visual clarity.}
	\Figlabel{lens_contamination_fraction}
\end{figure}

We use the entirety of these galaxies for our source sample, and all galaxies with Z\_B $ \leq 1.3 $ for our lens sample. In selecting lens-source pairs, we require that the source be in the background of the lens, separated by a minimum redshift of $\Delta z = 0.2$. This value is chosen to minimize the the fraction of galaxies identified as appropriate sources which in fact reside in the same bin as the lenses. To determine this fraction, we matched the CFHTLenS galaxy catalogue to publicly-available spectroscopic galaxy catalogues from the VIMOS VLT Deep Survey \citep[``VVDS''][]{LeVetGar05,GarLeGuz08} and the DEEP2 galaxy survey \citep{DavGuhKon07,NewCooDav13}, both of which overlap portions of the CFHTLenS. We then calculated the fractions of galaxies with sufficiently-large photometric redshifts which have spectroscopic redshifts within a given bin. As shown in \Figref{lens_contamination_fraction}, a buffer of $\Delta z = 0.2$ ensures that fewer than $1$ per cent of galaxies identified as appropriate sources in fact reside in the same bin as the lenses for lens redshift bins $z\sbr{mid} \leq 0.65$, and the benefit of increasing this value further is indistinguishable from noise.

\section{Application to Data}
\Seclabel{Testing}

In order to provide a proof-of-principle that our proposed method can be successfully used to measure the magnification signal from weak gravitational lensing, we test it by measuring the galaxy-galaxy lensing signal in the CFHTLenS. We choose the galaxy-galaxy lensing application as it is computationally straightforward to measure the shear signal at the same time as the magnification signal. As weak lensing shear is a well-studied and robust method for determining halo mass profiles with signal-to-noise of the same order of magnitude as weak lensing magnification, it provides an ideal comparison.

One drawback in using shear to model halo mass profiles is that the shear signal is degenerate to the addition or subtraction of a sheet of constant projected mass density to the field of view. Magnification is not subject to this degeneracy, which leads to the result that a comparison with shear alone cannot be used to determine whether a flat offset in the magnification signal is due to a bias in the measurement or to the presence of an actual over- or underdensity in the region of interest. See the discussion on the potential for lens-source correlations or anti-correlations to cause such a bias in the magnification signal in \Secref{determining_source_density}.

In \Secref{modelling_lensing_signal}, we discuss the model we use for the expected form of the lensing signal, and in \Secref{test_results}, we present the results of our tests.

\subsection{Modelling Lensing Signal}
\Seclabel{modelling_lensing_signal}

To fit our measured lensing signals, we assume that the underlying mass distribution can be described by a profile which is on average radial, and which can be described with relatively few parameters. The most common model used for this purpose is known as the halo model \citep{Sel00,ManTasSel05}, which has two free parameters: the halo mass of the galaxies in the sample, and the satellite fraction, with other parameters directly fixed through a physical model. For this paper, however, we choose to use a different model, which has one additional parameter to characterize the surrounding environment of galaxies in the sample. This allows us to better investigate how magnification might aid in breaking degeneracies between separate factors which characterize the environment of galaxies.

Our model uses four free parameters: $M\sbr{1h}$, the total one-halo mass of the lens; $M\sbr{gr}$, the mean mass of the group or cluster a typical satellite lens resides in; $f\sbr{sat}$, the fraction of lenses which are satellites of a group; and $\kappa\sbr{offset}$, a flat offset to the measured $\kappa$ due to observational biases. Other relevant parameters are fixed directly from observations or with fitting functions. We model the galaxies in each bin as having their mass dominated by a dark matter halo with a truncated \citet[]{NavFreWhi97} profile, using the model of \citet{BalMarOgu09} (hereafter a ``tNFW'' profile). We leave the mass of this profile as a free parameter, determine the concentration through the fitting function of \citet{NetGaoBet07}, and set the truncation radius equal to double the virial radius.

Of these galaxies, we assume that a fraction $f\sbr{sat}$ are satellites of a group, where $f\sbr{sat}$ and $M\sbr{gr}$, the total mass of the group, are free parameters. We assume the group's total mass can also be modelled with a tNFW profile, and we determine its concentration and truncation radius in the same manner as for the galaxy's halo. We assume that the distribution of satellites relative to group centre can be described by a scaled tNFW density profile with concentration $c=2.5$, which is consistent with the results of \citet{McGBalBow09}. The details of calculating the contribution of the group environment to the lensing signal can be found in \citet{GilHudErb13}, which uses the same model for shear except with a fixed $f\sbr{sat}$. For computational simplicity, we do not include any contributions from large-scale structure to the lensing signal (the so-called ``two-halo term'') at this stage of analysis, but we plan to do so in future work.

This model results in a mean mass profile for a given set of parameters $M\sbr{1h}$, $M\sbr{gr}$, $f\sbr{sat}$. To determine the shear and magnification lensing signals, we project this mass profile along the line of sight\footnote{For computational simplicity, we neglect the presence of structure correlated to the galaxy and group, the so-called ``two-halo'' term in the lensing signal.} to get the projected overdensity $\Sigma\left( \vec{R} \right)$. The expected shear and convergence signals can then be obtained through:
\begin{equation}
	\expec{ \gamma\sbr{t} } = \frac{\overline{\Sigma\left( <R \right)}-\overline{\Sigma\left( R \right)}}{\Sigma\sbr{crit}}
\end{equation}
and
\begin{equation}
	\expec{ \kappa } = \frac{\overline{\Sigma\left( R \right)}}{\Sigma\sbr{crit}}\mathrm{,}
\end{equation}
where $\overline{\Sigma\left( R \right)}$ is the mean projected density at a distance $R$ from the galaxy, $\overline{\Sigma\left( <R \right)}$ is the mean projected density contained within a distance $R$ of the galaxy, and $\Sigma\sbr{crit}$ is the critical projected density, using the same definition as \Eqref{Sigma_crit_def}. 

Finally, we assume that the correlation and anti-correlation between the number counts of lenses and sources across the field of view due to observational effects will have the net effect of shifting the measured magnification signal by a constant value, which we leave here as a free parameter, $\kappa\sbr{offset}$. While the other three free parameters affect both the measured shear and magnification signals, $\kappa\sbr{offset}$ only has an effect on the magnification signal.

We determine the best-fit set of parameters and their associated errors for each bin of interest through a three-phase Markov chain Monte Carlo (MCMC) algorithm. In the first and second phases, we follow the Metropolis-Hastings algorithm \citep{MetRosRos53}, using a $\chisq$ comparison of the model lensing signal with the measured values to determine the acceptance ratio. The first phase is a ``burn-in'' period, in which the algorithm is given time to reach the region with the lowest $\chisq$. In the second phase, we periodically record the points arrived at by the algorithm. The standard deviations in these values represent the $1\sigma$ errors in the best-fit model parameters. In the final phase, we apply simulated annealing, successively decreasing the size of steps and the acceptance ratio, to arrive at the best-fit set of parameters.

For each bin of interest, we fit the data in three manners: To magnification information only, to shear information only, and to both magnification and shear information.

\subsection{Resulting Measurements of Magnification Signal}
\Seclabel{test_results}

\begin{figure*}
	\includegraphics[scale=0.9]{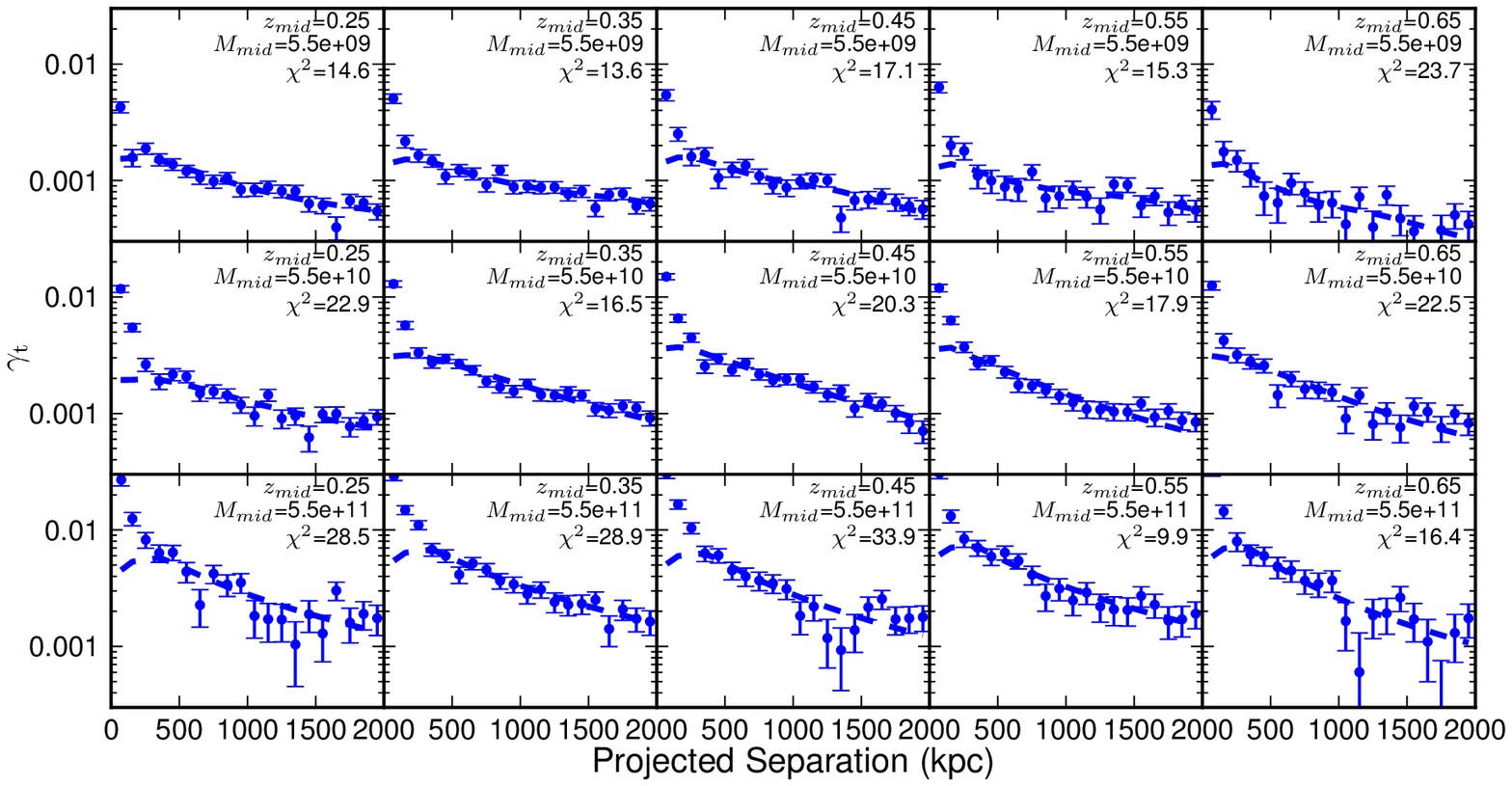}
	\includegraphics[scale=0.9]{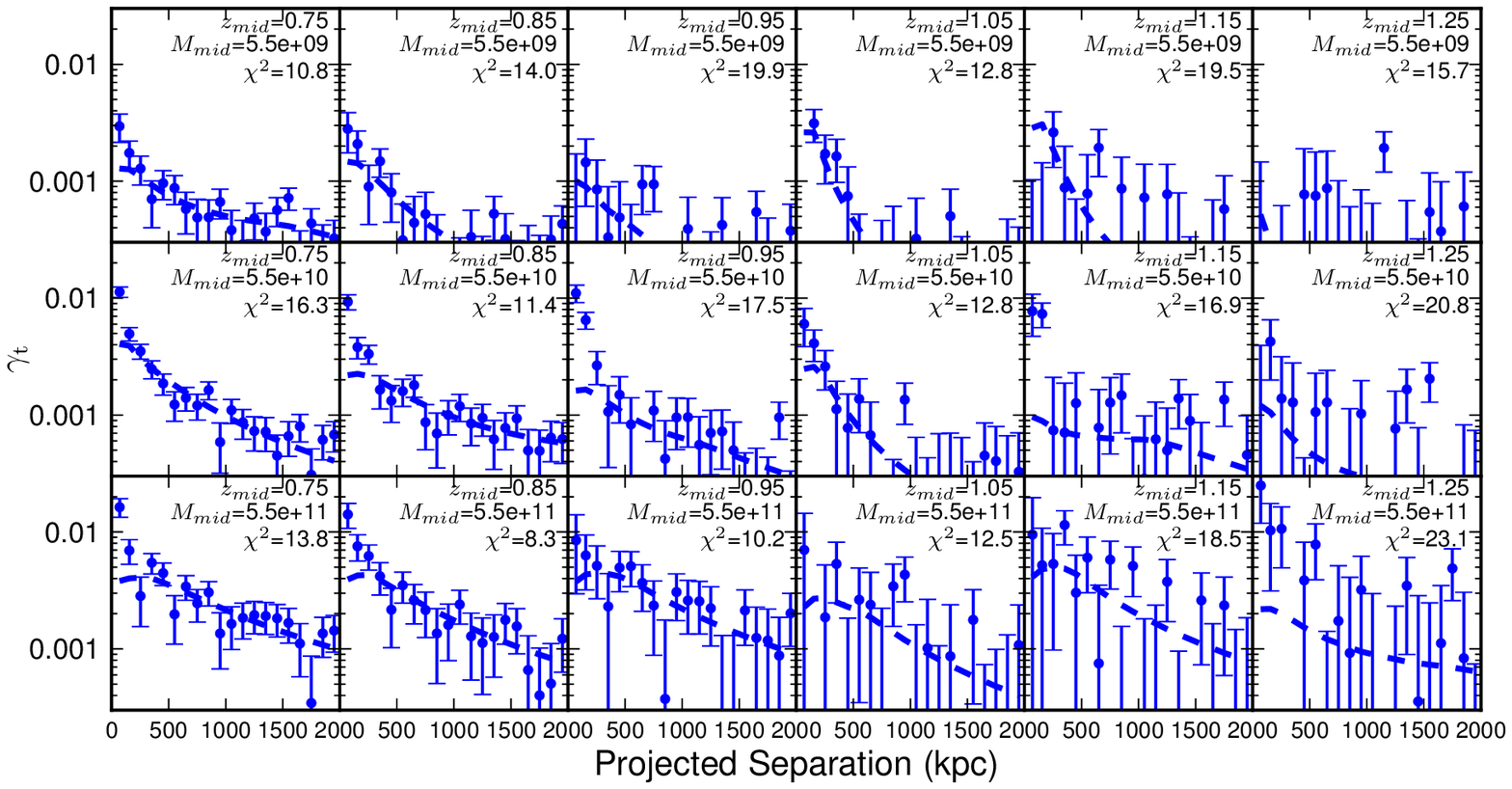}
	\caption[Measured shear signal $\gamma\sbr{t}$ overlaid with best fit model from magnification data alone.]{Measured shear signal $\gamma$ (points and errorbars) for different lens redshift and mass bins. Also shown is the best-fit model using shear data alone (dashed line). The top panel displays low-redshift bins, and the bottom panel displays high-redshift bins. The $\chisq$ values represent the $\chisq$ for a comparison of the shear data with the shear-only fit (15 dof.).}
	\Figlabel{gamma_with_models}
\end{figure*}

In order to test that our model is performing as expected, we first perform a measurement of tangential shear in various lens stellar mass and redshift bins and test fitting out model to the resulting signal. The results of this test are shown in \Figref{gamma_with_models}, where we present the shear signal in terms of the reduced tangential shear $\gamma\sbr{t}$ against angular separation for each lens bin as well as the predicted shear for the best-fit model. In each bin, the model passes a $\chisq$ test in comparison with the data, as does the mean $\chisq$ of $17.6$ for 15 degrees of freedom and 15 bins. Our model thus appears to provide a good fit to the shear data.

\begin{figure*}
	\includegraphics[scale=0.9]{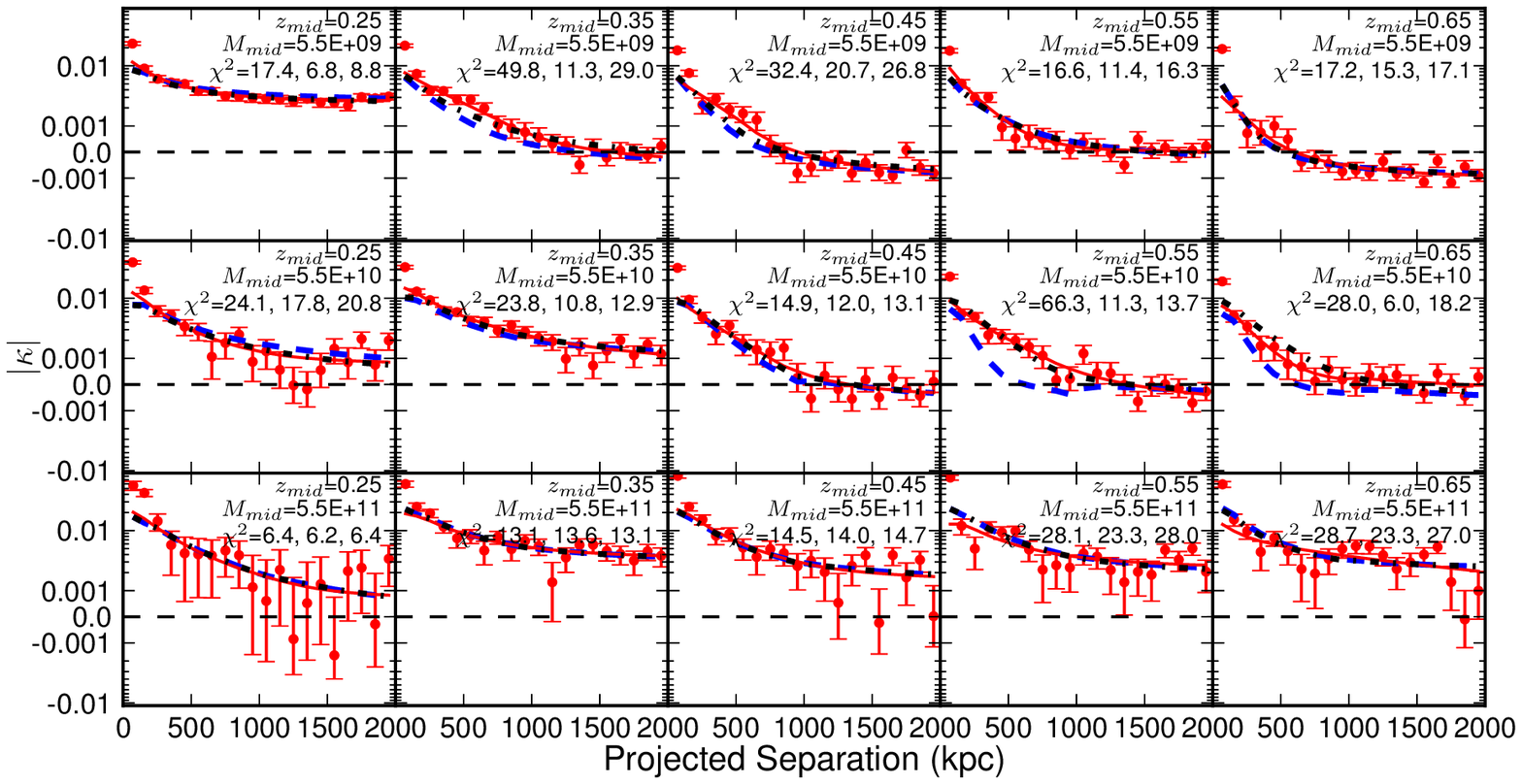}
	\includegraphics[scale=0.9]{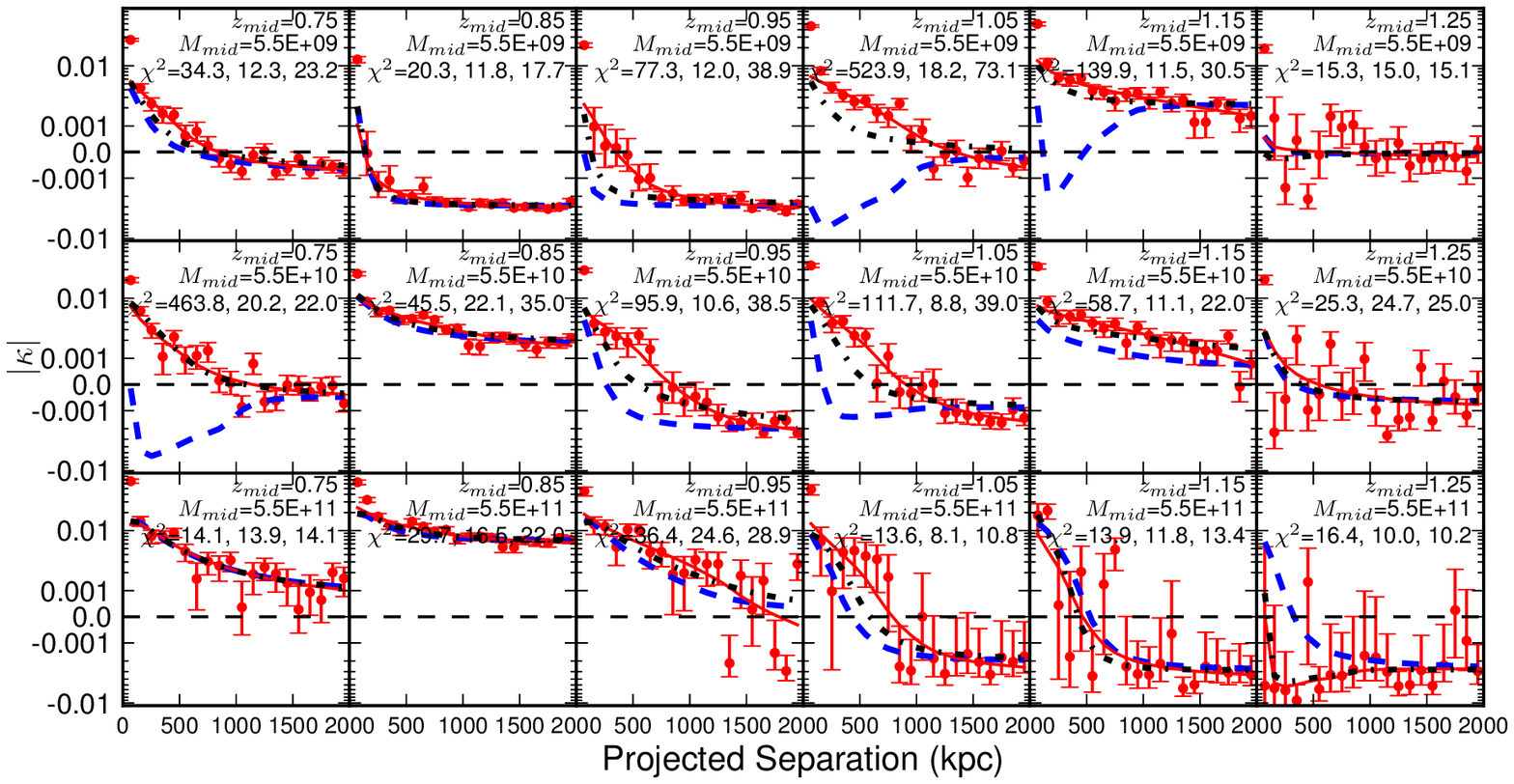}
	\caption[Measured magnification signal $\kappa$ overlaid with best fit model from shear data alone.]{Measured magnification signal $\kappa$ (points and errorbars) for different lens redshift and mass bins. Also shown are the best-fit models using shear data alone (blue dashed line) using magnification data alone (red solid line), and for the combination of the shear and magnification data (black dot-dashed line). The top panel displays low-redshift bins, and the bottom panel displays high-redshift bins. As the fit using shear data alone does not constrain $\kappa\sbr{offset}$, the fitted value from using magnification alone is used for that model. The $\chisq$ values represent the $\chisq$ for a comparison of the magnification data with the magnification-only fit (14 dof.), the shear-only fit (17 dof.), and the combined fit (14 dof.), respectively.}
	\Figlabel{kappa_with_models}
\end{figure*}

In \Figref{kappa_with_models}, we show the measured magnification signal for the same lens stellar mass and redshift bins as in \Figref{gamma_with_models}, presented in terms of the convergence $\kappa$, which is related to the measured magnification through \Eqref{kappa_estimate}. Also plotted on it are curves representing the predicted $\kappa$ from the best-fit model which uses only magnification data, the best-fit model which uses only shear data (for which $\gamma\sbr{t}$ for each bin is plotted in \Figref{gamma_with_models}), and the best-fit to the combination of the shear and magnification datasets. As the fit using shear data alone does not constrain $\kappa\sbr{offset}$, the fitted value from using magnification alone is used for that model. The data generally shows the expected form and is well-fit by our model, though the errors are typically larger than for the shear measurements. Notice that in some bins the signal converges to a positive value at large angular separation between lenses and sources, and in others it converges to a negative value. This is most likely an effect of variable seeing and other position-dependent effects leading to correlation or anti-correlation between lenses and sources, as we discussed in \Secref{determining_source_density}, and which is the reasoning behind our use of the $\kappa\sbr{offset}$ model term.

We note that in many cases, the innermost radial bins show very high magnification signals compared to the models. We believe this to be most likely due to the obscuration of background galaxies by dust in the regions of lens galaxies (particularly those which reside within groups or clusters). Dust can most easily obscure the faintest magnitude galaxies, and it is at the faintest magnitudes that the slope of the luminosity function is shallowest (as can be seen in \Figref{ex_count_and_alpha}). At these magnitudes, $\alpha(m)-1$ is negative, and so the the obscuration of a galaxy would result in a spurious increase in the magnification signal. As we know that the innermost radial bins are likely to be problematic in this manner, we leave them out of our fitting procedure and calculation of $\chisq$ values. Methods have been proposed to handle dust extinction (eg. \citealt{MScrFuk10}, which finds that magnification is typically a stronger effect than dust extinction by a factor of $\sim 3$ in the {\it V} band, and \citealt{BauGazMar14}), but as this paper is intended to serve as a first demonstration for the magnification methodology proposed here, we leave the full application of these for future study.

In the low redshift bins, the model which is best fit to only shear data generally also does a good job at fitting the magnification data. It fails a $\chisq$ test at the $5$ per cent threshold for only five of the fifteen low-redshift bins, and the fit to the combined dataset fails for only two of these bins: the $9 < \log{M\sbr{lens}/\msun} < 10$ and $0.3 < z < 0.4$ bin, and the $11 < \log{M\sbr{lens}/\msun} < 12$ and $0.6 < z < 0.7$ bin. This failure ratio of $13$ per cent for the combined fit does imply that something beyond statistical error is to blame here, but this is not surprising given the simple model we use in this paper.

In the higher redshift bins, the best-fit model to shear-only does a much poorer job at fitting the magnification signal. This is not surprising, as at these redshifts the signal-to-noise of the shear measurements is much lower, and so noise in the shear signal can propagate through into noise into the best-fit measurements, resulting in a high $\chisq$ value when it is compared with the magnification data. The fit to the combined dataset does a better job at fitting to both, though it still fails at the $5$ per cent threshold for seven of the eighteen high-redshift bins. It remains for future investigation to determine whether this is due to issues with the shear signal, the magnification signal, the modelling process, or a combination of these factors.

Overall, the strong fit of our model implies that our method is successfully recovering a real magnification signal. This magnification signal may, however, still have biases in it due to effects we haven't fully corrected for, such as dust obscuration or contamination of the source sample. Thus at this stage, the specific values fit by our model may still be biased. Given the strong agreement with the shear-signal, these biases are likely at a small level, however, and we can still gain insight into how magnification data can be used to improve the errors on fits to various parameters.

\begin{figure*}
	\includegraphics[scale=0.65]{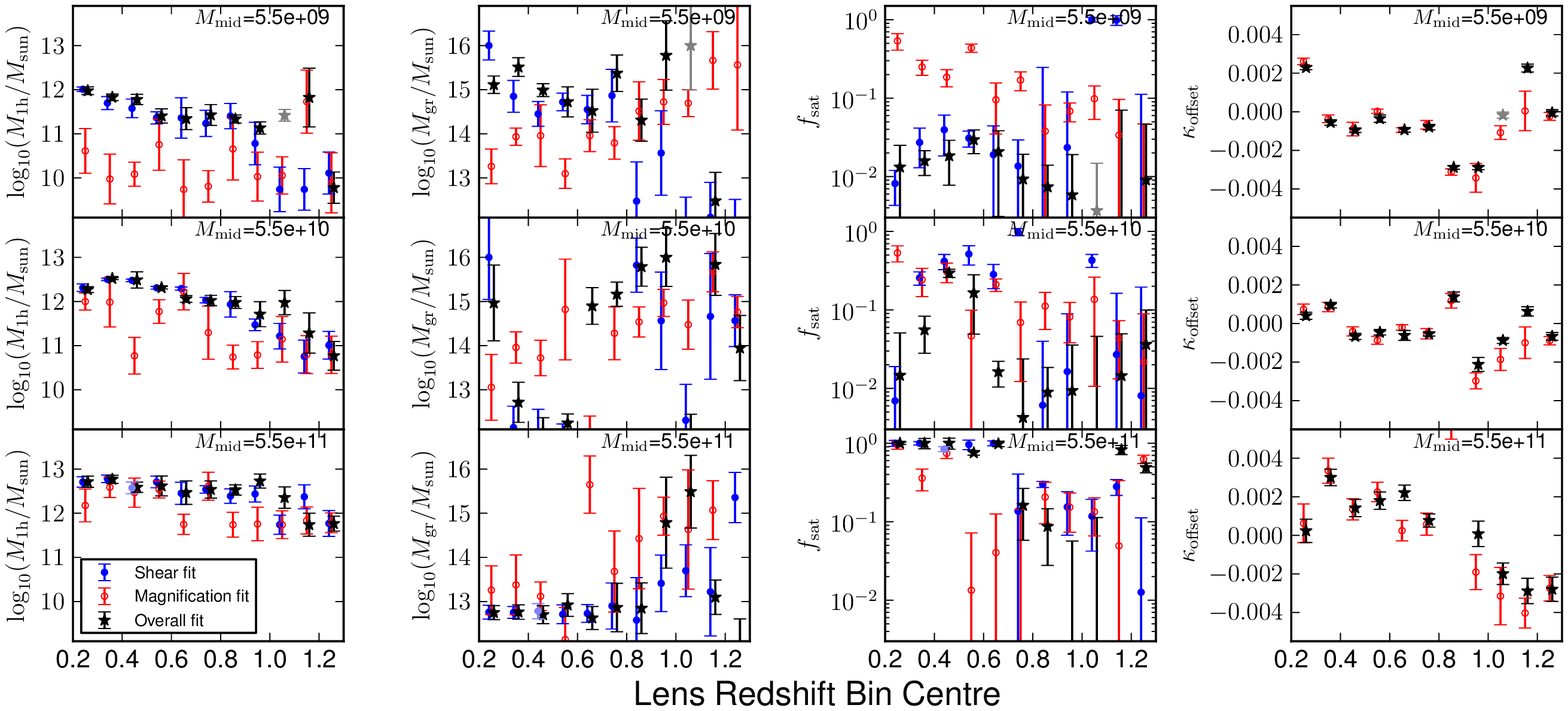}
	\caption[Best-fit values of our model parameters for various mass and redshift bins]{Best-fit values and errors of our model parameters, for various lens redshift and mass bins. From left to right, the columns show the parameters: One-halo mass, Group mass, Satellite fraction, and $\kappa\sbr{offset}$. The best-fit values using shear data alone are shown in blue with solid circle markers, values from magnification data alone are shown in red with open circle markers, and values from the combined data are shown in black with star markers. As shear data alone cannot constrain $\kappa\sbr{offset}$, the shear-only fit is not shown in that panel. Bins for which the $\chisq$ value indicates a poor fit ($>30$ for shear- and magnification-only fits, $>50$ for the fits to both) are not shown, as the best-fit parameters for these bins are not meaningful.}
	\Figlabel{lensing_model_fits}
\end{figure*}

This can be seen in \Figref{lensing_model_fits}, where we show the best-fit parameters for our model fits and their associated errors for all lens stellar mass and redshift bins tested. This figure shows the fits to shear data alone, magnification data alone, and to the combined dataset. As shown in \Figref{lens_contamination_fraction}, the contamination of the source sample with galaxies in the lens plane which were spuriously misidentified as sources becomes a significant issue at redshifts greater than $\sim 0.7$, and so the fits including magnification data at these redshifts may be less reliable. The fitted $M\sbr{1h}$ and $M\sbr{gr}$ parameters from magnification alone seem to be in general agreement with those from shear alone. It is in the $f\sbr{sat}$ parameter that the two fits differ, with the magnification-only fit regularly favouring a larger $f\sbr{sat}$.

Interestingly, some of the fitted $\kappa\sbr{offset}$ values show a significant change between the magnification-only fit and the fit to the combined dataset. This seems to occur when the best-fit group mass parameter $M\sbr{gr}$ for magnification alone was spuriously high but was corrected to a more-reasonable value when complemented with the shear data. There are also bins in which the shear-only fit appears to show a spuriously high or low, but in which the fit to the combined dataset is much more reasonable. This clearly shows the potential benefit of combining shear and magnification data.

The trend observed in the best-fit values for the one-halo mass parameter $M\sbr{1h}$ at large redshift is potentially worrisome - the combined fit seems to consistently favour very low masses, often lower than either the shear or magnification data alone favours. This is possibly due to the fact that we are not fitting to the innermost radial bins, which means we are losing a significant amount of information relevant to the one-halo mass fit. This could also potentially be due to contamination of the source sample in high lens-redshift bins leading to spuriously low magnification signals. We discuss this possibility further in \Secref{Estimating_contamination}.

\begin{figure*}
	\includegraphics[scale=0.65]{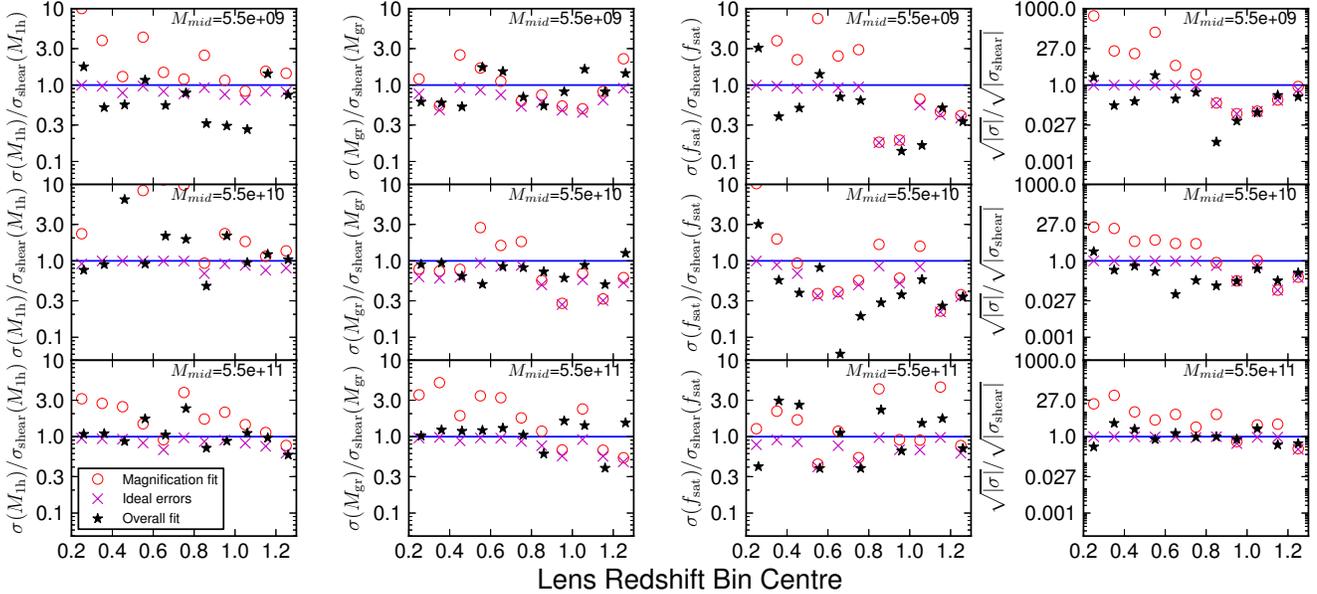}
	\caption[Relative our model parameters for various mass and redshift bins]{The ratios of errors in our model parameters to the errors obtained from using shear information alone, for various lens mass and redshift bins. From left to right, the columns show the parameters: One-halo mass, Group mass, Satellite fraction, and the square-root of the determinant of the covariance matrix, which is representative of the total volume of the confidence regions. The red open circles show the ratios of errors from magnification alone to the errors from shear alone, the black stars show the ratios of errors from the combination of shear and magnification to shear alone, and the magenta crosses show the ratio of the theoretical combination of errors from magnification and shear to the errors from shear alone. Note that due to the multi-parameter nature of the fit and the fact that the best-fit values from shear and magnification alone sometimes disagree, the errors from the combined dataset are occasionally greater than for shear alone.}
	\Figlabel{lensing_model_errors}
\end{figure*}

In \Figref{lensing_model_errors}, we show the ratios of the calculated errors on our fitted parameters to the errors obtained when shear information alone is used. Both the ratios of the magnification-only fit and the combined dataset fit are shown, as well as a calculation of the theoretical ratio, where the theoretical combined error is:
\begin{equation}
	\sigma\sbr{th} = \left( \sigma\sbr{shear}^{-2} + \sigma\sbr{mag}^{-2} \right)^{-1/2}\mathrm{,}
\end{equation}
which is not necessarily valid for a fit of more than one parameter, depending on the relative degeneracies of the two fits.

This plot shows that when used alone, magnification typically does a poorer job at constraining the model parameters at low redshift. However, as expected it is more beneficial at high redshift, particularly for the parameters relating to the host groups of the galaxies in the sample, $M\sbr{gr}$ and $f\sbr{sat}$. Interestingly, the errors of the fit to the combined dataset do not generally follow the theoretical calculations, in some cases being even smaller, and in some cases being even larger than in the shear-only case. This suggests that the situation is more complicated than the theoretical calculation assumes.

What is likely going on is a combination of two factors. The shear and magnification signals and the fits to them have different degeneracies, and the two fits disagree on the best-fit set of parameters in some scenarios. The former factor can result in the fit to the combined dataset having significantly smaller errors than the fit to either alone, as the combination of the two datasets breaks the degeneracies that are present in either alone, allowing much greater precision. The latter factor can result in the fit to the combined dataset showing larger errors as the different regions preferred by each dataset will spread out the most-reasonable region in the fit to the combined dataset. The result is that when the regions preferred by the individual fits overlap, the fit to the combined dataset is much smaller, while when they do not overlap, it can be larger.

Notably, at high redshifts and low lens stellar masses, the shear signal approaches zero and isn't reliably detectable, as can be seen in the bottom panel of \Figref{gamma_with_models}. This results in the best-fit model parameters being unreliable when shear alone is used. However, as can be seen in the bottom panel of \Figref{kappa_with_models}, the magnification signal can in fact still be measured for these bins, and so the best-fit model parameters from using only magnification are likely to be more reliable.

\begin{figure*}
	\includegraphics[scale=0.45]{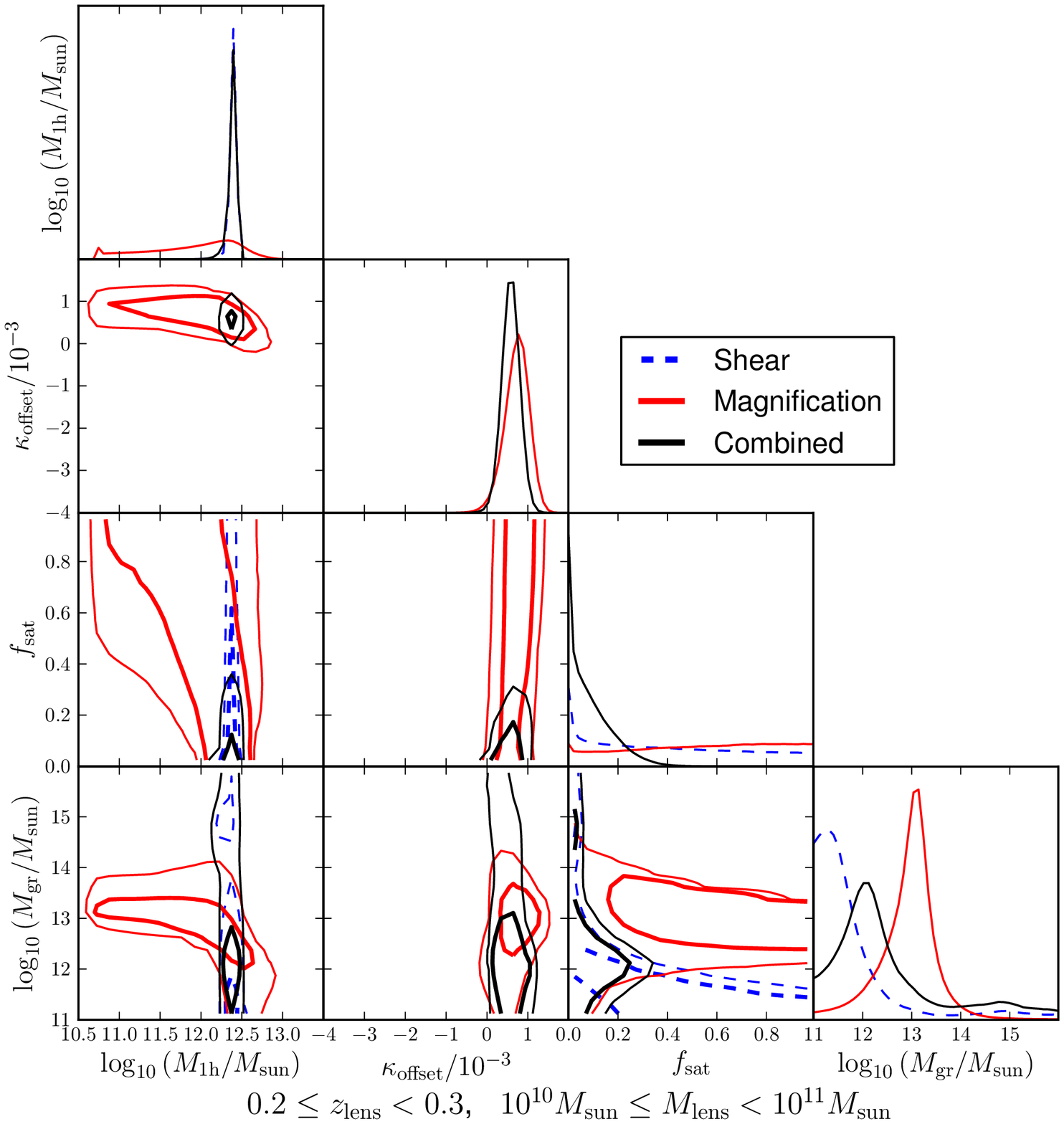}
	\includegraphics[scale=0.45]{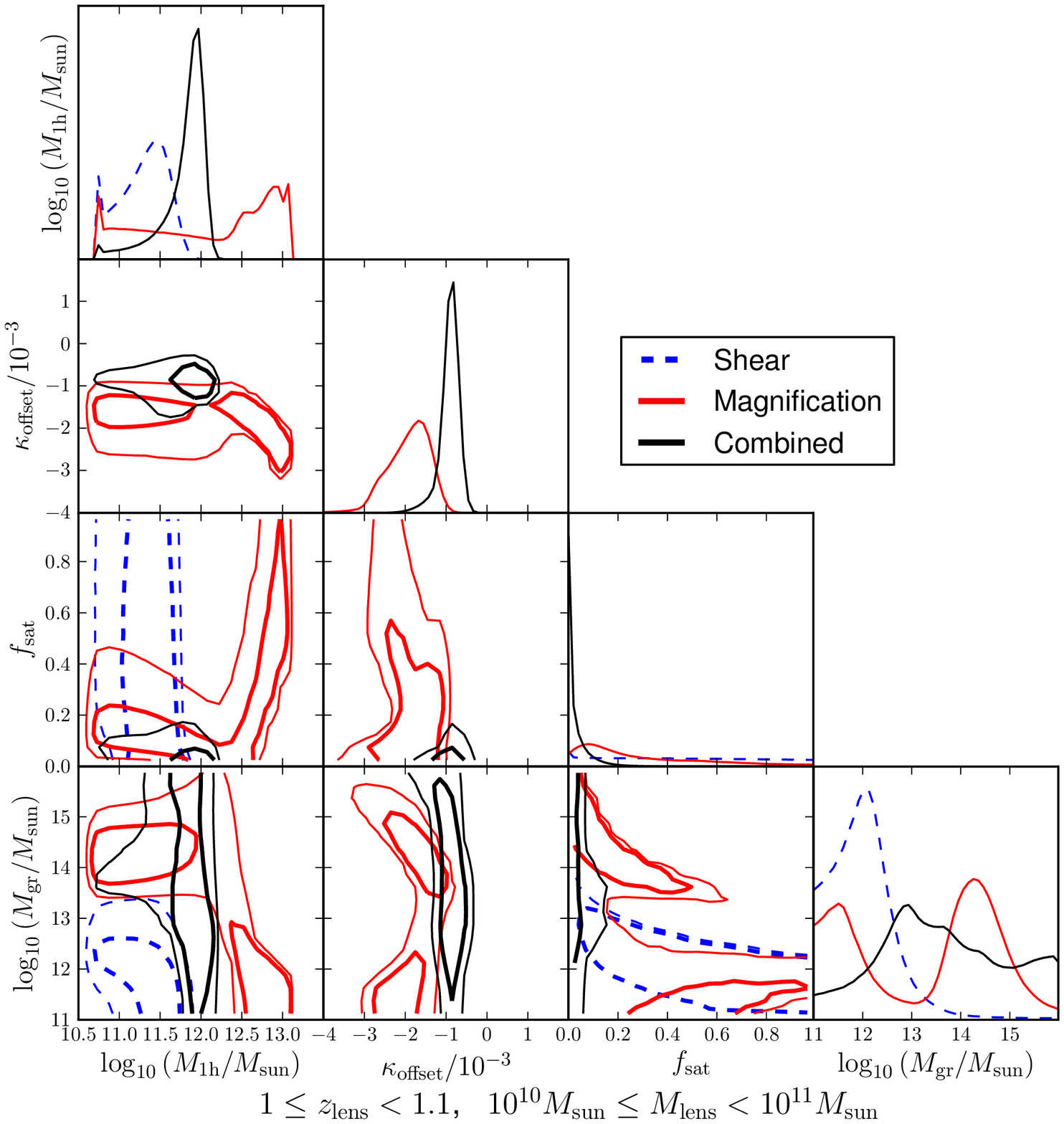}
	\caption[Covariance between fitted parameters for the z $\sim 0.25$, m  $\sim 10^{10.5}$ and z $\sim 1.05$, m  $\sim 10^{10.5}$ bins.]{Confidence regions of fitted parameters for shear-only, magnification-only, and combined fits to the $0.2 \leq z < 0.3$, $10^{10} \msun \leq M\sbr{lens} < 10^{11} \msun$ bin (left) and the $1 \leq z < 1.1$, $10^{10} \msun \leq M\sbr{lens} < 10^{11} \msun$ bin (right). The thick and thin contours show 95\% and 68\% confidence regions respectively. Note that the shear-only fits aren't shown for the panels which include the $\kappa\sbr{offset}$ parameter, as this parameter is unconstrained by the shear signal.}
	\Figlabel{fit_confidence_regions}
\end{figure*}

In order to help illustrate the possible benefits of magnification data, \Figref{fit_confidence_regions} shows the confidence regions for the fitted parameters to two lens bins, selected to illustrate cases in which magnification is particularly useful. These show that magnification provides the most benefit in the $M\sbr{gr}-f\sbr{sat}$ parameter plane. Here, the shear signal is only able to provide an upper boundary to the combination of $M\sbr{gr}$ and $f\sbr{sat}$. The magnification signal on its own is not much better, also covering larger regions of parameter space. However, the intersections of these confidence regions are relatively small, and so the fit to the combined dataset has a much tighter confidence region.

It remains for future investigation to determine how much benefit magnification data will provide when all relevant biases are taken into account. The fact that even for this preliminary analysis, magnification occasionally provides large benefits in reducing the errors in fitted parameters is strongly encouraging and motivates further investigation to determine whether this will remain the case in a proper treatment.

\section{Discussion}
\Seclabel{Discussion}

In \Secref{method_comparison}, we compare our method to the optimally-weighted correlation function method, which has been used in past analyses. In \Secref{mass_mapping_implementation}, we present a proposed implementation of our methodology for mass-mapping. In \Secref{Estimating_contamination}, we present an estimate of the possible impact of source sample contamination on our results. In \Secref{Field_weighting_effects}, we test and discuss the impact of our choice to weight our measurement of the unmagnified source density on a per-field basis. Finally, in \Secref{Further_Research}, we discuss further possible work.

\subsection{Comparison to Other Methods}
\Seclabel{method_comparison}

When applied to galaxy-galaxy lensing, the results of this methodology are nearly equivalent to those obtained through an optimally-weighted cross-correlation function estimator. This method was first proposed by \citet{MBar02}, and in typical implementations \citep[eg. ][]{ScrMRic05,HilPieErb09,ForHilVan12} uses a cross-correlation function in which all observed sources are weighted by $\alpha(m)-1$. The measured correlation function can then be compared to the predicted correlation functions from mass models. To date, it has not been stated in the literature how this correlation function relates to the magnification, and so we investigate this here.

The angular cross-correlation represents the excess number density of galaxies in one sample in angular bins around galaxies in another sample. For scenarios in which only magnification is expected to influence the correlation between lens and source positions, which generally requires the two samples to be well-separated in redshift, the expected cross-correlation between lenses and sources in a given angular and magnitude bin would be
\begin{equation}
	\expec{w(\theta,m)} = \frac{\expec{N(\theta,m)}}{N\sbr{0}(m)} - 1 = \mu(\theta)^{\alpha(m)-1} - 1\mathrm{,}
\end{equation}
where $N\sbr{0}(m)$ is the unmagnified number count of source galaxies at a given magnitude and $\expec{N(\theta,m)}$ is the expected magnified number counts of galaxies at a given magnitude in a given angular bin relative to the lens sample, and using \Eqref{mu_definition} to identify this ratio as corresponding to $\mu(\theta)^{\alpha(m)-1}$.

Approaches by eg. \citet{ScrMRic05}, \citet{HilPieErb09}, and \citet{ForHilVan12} calculate the following optimally-weighted cross-correlation function, using a weighted version of the \citet{LanSza93} estimator,
\begin{equation}
	w\spr{opt}\sbr{LS} = \frac{\rm LS^{\alpha-1} - RS^{\alpha-1} - \expec{\alpha-1}LR  }{\rm RR} + \expec{\alpha-1}\mathrm{,}
	\Eqlabel{LS_cross_corr_opt_estimator}
\end{equation}
where S stands for \textit{sources}, L stands for \textit{lenses}, and R stands for the \textit{random} catalogue, which is at least 10 times as large as the lens and source catalogues. All sources are weighted by their $\alpha(m)-1$ value, $\expec{\alpha-1}$ is the mean value of $\alpha(m)-1$ across the entire sample, and all pair counts are normalized by the product of the total numbers of galaxies in each catalogue.

This can be more easily compared to our method by instead presenting the cross-correlation function using the \citet{DavPee83} estimator, which converges to the same result as the \citet{LanSza93} estimator at high number counts. The optimally-weighted cross-correlation function with this estimator is:
\begin{equation}
	w\spr{opt}\sbr{DP} = \frac{\rm LS^{\alpha-1} }{\rm LR} - \expec{\alpha-1}\mathrm{,}
\end{equation}
using the same notation as in \Eqref{LS_cross_corr_opt_estimator}. The optimally-weighted correlation can therefore be seen as a measurement of
\begin{equation}
	w\spr{opt} = \frac{\sum_{j \in {\rm annu}} \left[\alpha(m_j)-1\right]}{N\sbr{0,{\rm annu}}} -
		\frac{\sum_{j \in {\rm samp}} \left[\alpha(m_j)-1\right]}{N\sbr{0,{\rm samp}}},
\end{equation}
where $j \in {\rm samp}$ represents the indices of all source galaxies in the sample, $j \in {\rm annu}$ represents only those source galaxies which lie in the appropriate annulus behind any lens in the sample, $N\sbr{0,{\rm samp}}$ is the total number of source galaxies in the sample, and $N\sbr{0,{\rm annu}}$ is the expected number of source galaxies lying in the annuli in the presence of no magnification. If we convert this sum to an integral in the same fashion as in \Eqref{n_of_m_differential_form}, we get:
\begin{equation}
	w\spr{opt} = \frac{\int_a^b n(m)\left[\alpha(m)-1\right]dm - \int_a^b n\sbr{0}(m)\left[\alpha(m)-1\right]dm}{\int_a^b n\sbr{0}(m)dm}\mathrm{.}
\end{equation}
Compare this to the expansion of \Eqref{mu_hat_integral_form}:
\begin{align*}
  \hat{\mu} = &\;1 + \frac{\int_{a}^{b} n(m)\left[\alpha(m)-1\right] dm - \int_{a}^{b} n\sbr{0}(m)\left[\alpha(m)-1\right] dm}{\int_{a}^{b}n_{0}(m)\left[\alpha(m)-1\right]^2 dm} \mathrm{.} \numberthis
  \Eqlabel{mu_hat_expanded_form}
\end{align*}

In both equations, the number count of galaxies observed within an annulus of interest only appears within one term, $\int_a^b n(m)\left[\alpha(m_j)-1\right]dm$. The methods thus have equivalent statistical power, and it is possible to convert between their measurements through:
\begin{equation}
	\hat{\mu} = 1 + \frac{ w\spr{opt}}{\expec{\left[\alpha-1\right]^2}}\mathrm{,}
	\Eqlabel{mu_hat_from_w}
\end{equation}
and
\begin{equation}
	w\spr{opt} = \expec{\left[\alpha-1\right]^2}(\hat{\mu} - 1)\mathrm{,}
	\Eqlabel{w_from_mu_hat}
\end{equation}
where
\begin{equation}
	\expec{\left[\alpha-1\right]^2} = \frac{\int_a^b n\sbr{0}(m)\left[\alpha(m)-1\right]^2dm}{\int_a^b n\sbr{0}(m)dm}\mathrm{.}
\end{equation}

This comparison allows the statistical error of $w\spr{opt}$ to be easily estimated without expending computational power on a bootstrap or jackknife approach. It can instead be estimated as:
\begin{equation}
	\stderr{w\spr{opt}} = \sqrt{ \frac{\expec{\left[\alpha-1\right]^2}}{N\sbr{0}A_{\rm samp}} }\mathrm{.}
	\Eqlabel{w_std_err}
\end{equation}
The derivation of this equation is presented in \Eqref{w_std_err_derivation}. This estimate relies on the assumption that background galaxies follow an ideal Poisson distribution. Since clustering is expected among these galaxies, their actual distribution will have a larger standard deviation than that of a Poisson distribution, and so this estimate will be biased low.

Computationally, both our method and the correlation function method scale as at best $O(N\sbr{l}\log{N\sbr{s}})$ with the number of lens galaxies $N\sbr{l}$ and number of source galaxies $N\sbr{s}$. This is due to the fact that they both include a step in which the source galaxies residing in a given annulus around lens galaxies must be determined, a process which can have this scaling at best. The difference comes in the coefficient of this term, which is dominated by the calculation of errors. Our method can calculate errors with only a few extra calculations per lens, as detailed in \Secref{galgal_implementation}. However, for the correlation function method, a bootstrap or jackknife method is typically used to determine the errors, which requires re-running the analysis on subsets of the data multiple times (eg. 50 times in the case of \citealt{ForHilVan12}).

The comparison we show here allows calculations of the magnification signal through a correlation function estimator to use \Eqref{w_std_err} as an estimate of the error. However, this estimate will be biased low due to the clustering of source galaxies. The algorithm we use to calculate a more accurate error estimate for our method (see Equations \ref{eq:mu_stderr_1} through \ref{eq:mu_stderr_3}) cannot be readily translated to a correlation function estimator, as it relies on the unmasked fractions of the annuli surrounding each lens in the sample, which are not calculated through a correlation-function-based approach. This results in our method being significantly faster than a correlation-function-based approach when accurate errors on the magnification estimate must be calculated, thanks to the fact that it does not have to resort to a computationally-expensive bootstrap or jackknife approach.

It's worth noting that many correlation function estimators, when measuring the function for large separations, bin objects into cells and determine the unmasked fraction of each cell \citep[eg.][]{ScrMRic05}. The correlation function between these cells is then calculated through a summation over pairs of cells. In this case, the error can be determined empirically through an analogous calculation to the one used for our method. However, such binning is only practical at large separations. Modifications can be made to the small-scale estimation to allow for error calculations, but this will require the calculation of the unmasked fraction of each annulus. If this is done, it is then no longer necessary to use a random catalogue to determine the overdensity within each annulus; since its area is known, its overdensity can be more efficiently calculated by simply comparing its density to the density of the full observation. At this point, the method is in fact more similar to our proposed method than to the traditional method of estimating a correlation function, lacking only the generalization done in our method to allow different redshifts for lenses. It is thus apparent that our proposed method is the most practical means to accurately determine the error in a measurement of a magnification signal.

Finally, our method also provides the advantage that it can handle overlapping lens and source samples. Correlation-function-based approaches require the lens and source samples to be well-separated in redshift space so that there is no expected clustering between the two samples. Our method has the benefit that each lens is treated individually, and so it is only necessary that each lens be well-separated from the subset of the source galaxy sample which we use to measure magnification. This results in the ability for lenses at the low-redshift end of the lens sample to use a greater amount of redshift space to measure the magnification signal, regardless of the fact that higher-redshift lenses in the sample cannot use this space. This will result in an increased signal-to-noise of the magnification measurement.

\subsection{Proposed Implementation for Mass Mapping}
\Seclabel{mass_mapping_implementation}

As we showed in \Secref{method_comparison}, our method has equivalent statistical power to cross-correlation function methods already in use when applied to galaxy-galaxy lensing. One important advantage this method has is that it is capable of providing an estimate of the mean magnification of an arbitrary region of space, not limited to a circular annulus, or even to a region of constant weight.

As the statistical power of magnification varies with the square-root of the area of the region measured, mass mapping with magnification alone must necessarily involve a trade-off between resolution and accuracy. The simplest approach to mass mapping would involve measuring the magnification, and thus projected overdensity, in a circular aperture around any point in the region of interest. While this method can be easily applied with the methodology outlined in the previous sections, it suffers from the drawback that the resulting mass map will show discontinuous behaviour whenever a source galaxy enters or exits the aperture.

This discontinuous behaviour can be avoided through instead applying a filter function which approaches zero at large separation, such as a two-dimensional Gaussian weight function. This also provides the benefit that source galaxies closer to the point of interest will have more influence on its measured signal than distant galaxies. Let us call this function $F(\theta)$.

The magnification at a given angular position $\vec{\theta}$ and redshift $z$ can then be estimated as
\begin{align*}
	\hat{\mu}_{kl} = &1 + \frac{1}{A\sbr{F}w(z+\Delta z)} \times \\ \numberthis
	&\left( \sum_{j} F\left( \left| \vec{\theta}'_j-\vec{\theta} \right| \right) \left[\alpha(m_{j})-1\right] - A\sbr{F}d(z+\Delta z) \right) \mathrm{,}
	\Eqlabel{mu_hat_mass_map_form}
\end{align*}
where $d(z)$ and $w(z)$ are defined as in \Secref{galgal_implementation}, $j$ represents a summation over all source galaxies at redshift $z+\Delta z$ or greater, with $\vec{\theta}'_j$ representing the angular coordinates of these galaxies, and
\begin{equation}
	A\sbr{F} = \iint F(\theta) d\Omega\mathrm{.}
\end{equation}
In practice, source galaxies sufficiently distant from the position of interest such that $F\left( \left| \vec{\theta}'_j-\vec{\theta} \right| \right) \rightarrow 0$ can be ignored in this calculation.

One notable issue with mass-mapping with magnification is the fact that position-dependent biases in the redshift determinations of source galaxies can result in a spurious magnification signal, as discussed in \Secref{determining_source_density}. In the scenario of galaxy-galaxy lensing, this effect is of large enough scale that it can be treated as a flat offset and included as a fitted parameter in the model, but this is not possible for mass mapping. The effect is somewhat reduced in mass-mapping, as the positions of lenses are not used, and so this effect should only appear in tomographic analyses in which a redshift cut-off is imposed on sources. This cut-off would allow for position-dependent effects, as the number of sources beyond the cut-off will depend on the errors in their redshift determinations, which is itself position-dependent. Care will thus have to be taken in such an analysis to account for this effect.

\subsection{Estimating Source Sample Contamination}
\Seclabel{Estimating_contamination}

As shown in \Figref{lens_contamination_fraction}, the fraction of galaxies whose photometric redshifts make them appear to be sources but which in fact lie within the lens plane can is of order $0.1$ per cent at low redshifts, and can be as high as $\sim 10$ per cent at high lens redshifts, even with a large buffer redshift between lenses and sources imposed. Galaxies in the lens plane will typically be correlated in position with each other due to their mutual gravitational attraction. Therefore, if galaxies in this plane are misidentified as sources, we would expect them to generally appear closer to lens galaxies than they would through random chance alone.

\begin{figure*}
	\includegraphics[scale=0.9]{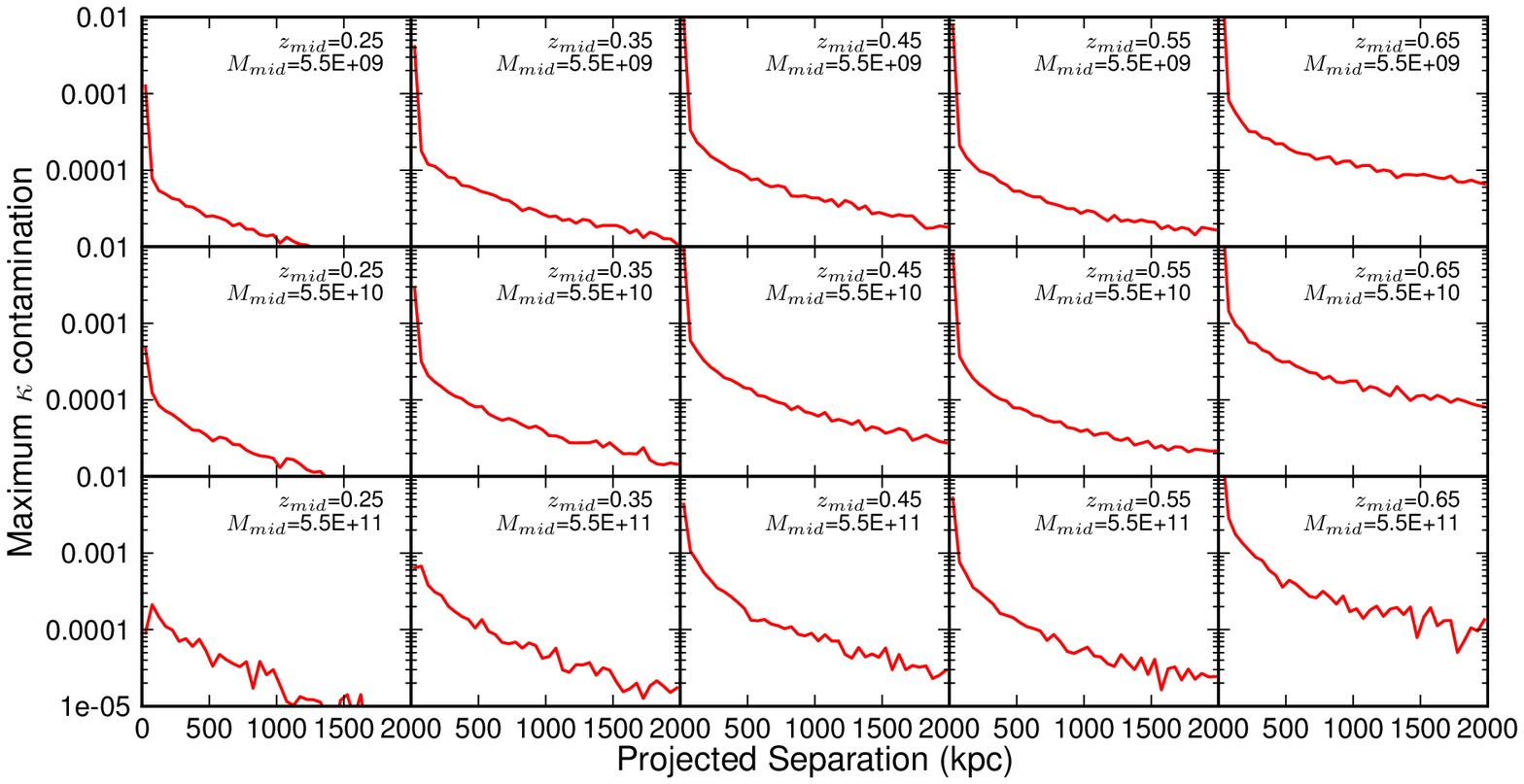}
	\includegraphics[scale=0.9]{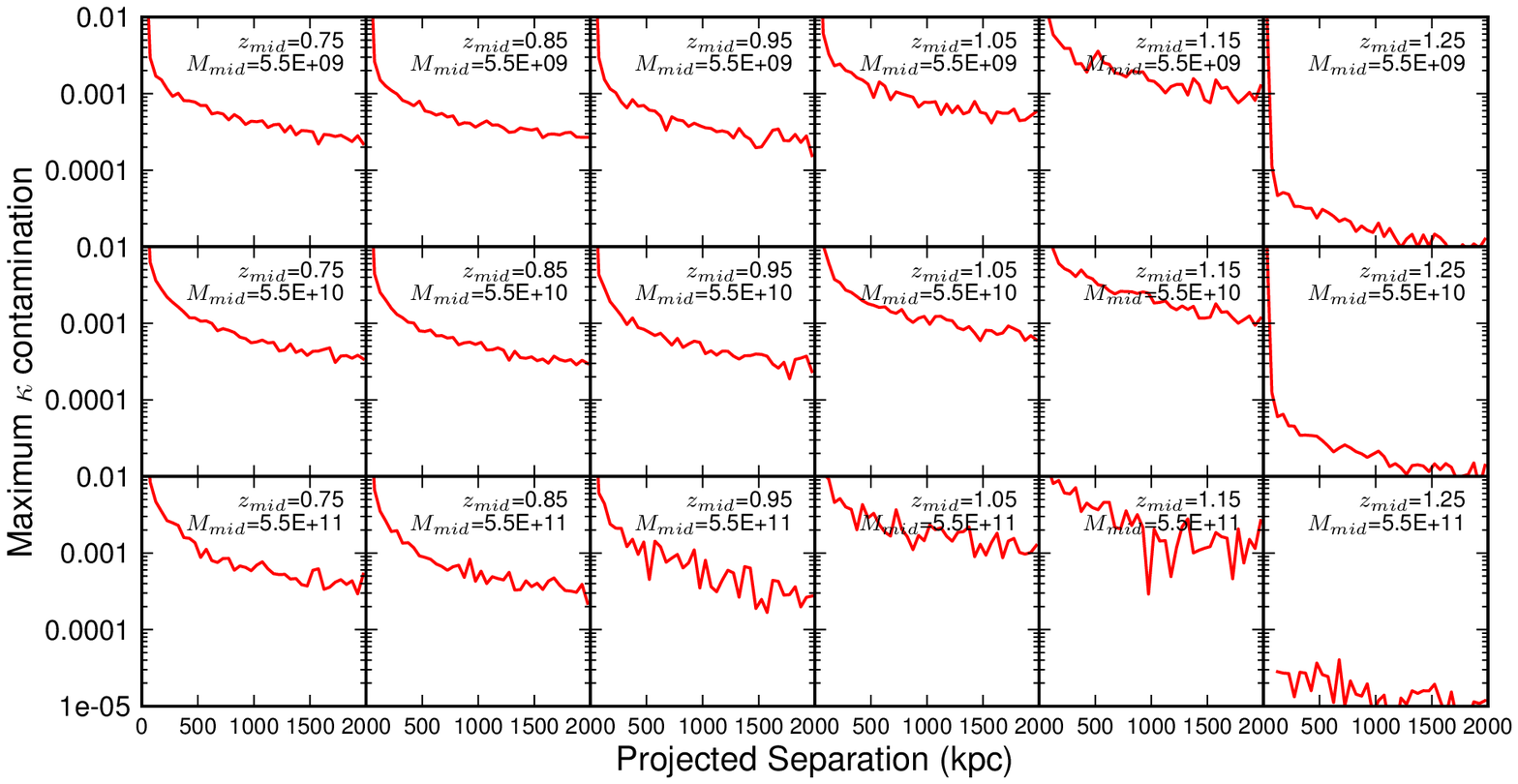}
	\caption[Maximum change in measured $\kappa$ due to contamination of source sample]{The maximum change in the measured $\kappa$ due to contamination of the source sample with lenses. Low-redshift bins are shown in the top panel, and high-redshift bins are shown in the bottom panel. This maximum change is calculated from the product of the correlation function shown in \Figref{lens_source_same_redshift_corr_funcs}, the contamination at the lens redshift, which is shown in \Figref{lens_contamination_fraction}, and the factor $0.25$, which represents a conservative upper bound on the slope of $\kappa/w$.}
	\Figlabel{kappa_contamination}
\end{figure*}

We can make an order-of-magnitude estimate of the impact this will have on our measured magnification signal through multiplying the correlation function of galaxies in the lens plane with the contamination and a factor representing the conversion from the correlation function to $\kappa$. For this conversion factor, we choose $0.25$, which would correspond to a mean $\left|\alpha(m)-1\right|$ of $\sim 0.5$. Typically, the mean is less than this, so this results in an upper bound on the contamination of $\kappa$. We plot the resulting predicted contamination for selected lens redshift and stellar mass bins in \Figref{kappa_contamination}. As can be seen through comparison with \Figref{kappa_with_models}, this value is typically negligible for $z < 0.6$, becoming comparable in magnitude to $\kappa$ in the $0.6 < z < 0.7$ bin. In higher redshift bins, the effect is even worse.

Further investigation of correcting for this effect is beyond the scope of this paper, but it is unlikely to be present significant difficulties. By virtue of the fact that the galaxies in the lens plane which are assigned photometric redshifts appropriate for sources, they likely have a similar magnitude distribution to legitimate source galaxies, and so correction for their differing magnitude distribution may not be necessary. As such, a simple calculation based on the correlation function of galaxies in the lens plane and the mean value of $\alpha-1$ for sources may be all that is needed.

\subsection{Effects of Field Weighting}
\Seclabel{Field_weighting_effects}

\begin{figure}
	\includegraphics[scale=0.45]{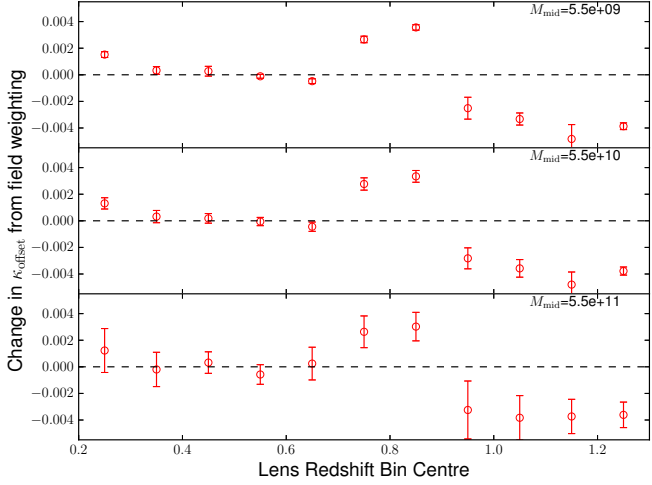}
	\caption[Demonstration of change in $\Sigma\sbr{offset}$ with field weighting.]{The change in the fitted values of $\kappa\sbr{offset}$ when weighting is applied on a per-field basis in the source-counting procedure. This weighting accounts in part for observational effects such as variable seeing which indirectly result in the number densities of lenses and sources being correlated or anti-correlated. The magnitude of the change shown here provides an estimate of the uncertainty in the value of $\kappa\sbr{offset}$ due to smaller-scale variations.}
	\Figlabel{kappa_offset_change_with_field_weighting}
\end{figure}

In \Secref{determining_source_density}, we discussed how correlations and anti-correlations of lenses and sources in regions of the sky can result in an effect which mimics a magnification signal. We chose to partially correct for this effect by applying weights on a field-by-field basis in determining the unmagnified source density. We can demonstrate the impact of this weighting through a comparison of the best-fit $\kappa\sbr{offset}$ parameters in the scenarios in which we do and do not apply field-by-field weighting, as shown in \Figref{kappa_offset_change_with_field_weighting}.

As expected, the application of weighting has the same effect on all lens stellar mass bins, but the effect varies with redshift. This is consistent with our hypothesis that variable seeing could result in varying amounts of scatter between redshift bins from field to field, as this effect would not be expected to be constant across all redshifts due to the complicated nature of determining photometric redshifts.

In general, the magnitude of change in $\kappa\sbr{offset}$ due to field-by-field weighting is $\lesssim 0.004$. Any remaining effect due to variable seeing on smaller scales is likely to result in a smaller change, and so we would expect the proper fitted values for $\kappa\sbr{offset}$ for all bins to be of smaller magnitude. From the fitted values shown in \Figref{lensing_model_fits}, it appears that the scatter in fitted values of $\kappa\sbr{offset}$ is comparable to this magnitude. Additionally, many bins of the same redshift but different stellar masses show significantly different fitting values of $\kappa\sbr{offset}$. This suggests that the fitted values for $\kappa\sbr{offset}$ are not sensitive to this effect alone, but are also affected by other aspects of the measured lensing signals. This is possibly due to the fact that the model we use does not include contributions for larger-scale structure, which is dominant at the same large radial separations that $\kappa\sbr{offset}$ is.

Alternatively, let us consider that this may be due not to an observational effect, but due to actual over- or underdensities in the regions of these lenses. The convergence maps generated by \citet{VanBenErb13} from shear data alone show that, when smoothed on scales of $2 \arcmin$ and correcting for the contribution of noise, the convergence has typical standard deviation of $\sim0.006$.\footnote{We determined this value through mass map data provided by the authors of \citet{VanBenErb13}. Under the assumption that the variance in the predicted convergence from their mass maps is the sum of the variances due to noise and from the actual variance in $\kappa$, we took the difference between the variances in $\kappa$ for their data maps and noise maps to estimate the variance, and thus standard deviation, in $\kappa$ due to the actual distribution of matter.} The typical area covered by all lenses within one of our stellar mass and redshift bins is of order $\sim10^5 \mathrm{arcmin}^2$. As this is roughly four orders of magnitude greater than the smoothing scale applied by \citet{VanBenErb13}, we would expect the standard deviation of $\kappa$ due to large scale structure within this region to drop by two orders of magnitude, to $\sim0.00006$. Clearly, this is too small to account for the scatter we observe in the fitted values of $\kappa\sbr{offset}$, and so they are more likely due to observational effects or fitting errors than cosmic variance.

\subsection{Further Research}
\Seclabel{Further_Research}

In our attempts to fit a mass profile to shear and magnification data simultaneously, we found several indications that our fit was not adequate, including poor $\chisq$ comparisons, extreme and unlikely values fit for certain bins, and inconsistent trends over changing lens redshift and stellar mass. It is possible that these issues might be resolved through the use of a different model mass profile, or through the inclusion of the effects of large-scale structure on the lensing signal. This will allow us to better determine the potential benefits of magnification in breaking degeneracies between parameters characterizing the environment surrounding galaxies in our samples.

As mentioned in \Secref{test_results}, dust obscuration can result in the measured magnification signal to be spuriously high if it is not corrected for. In this paper, we handled this effect through discarding the two innermost radial bins when we fit our mass profiles to the magnification data, as these are the bins which are likely to show the most significant contribution from dust. A proper analysis would require modelling the obscuration from dust, as done by eg. \citet{MScrFuk10}. This will be necessary before firm conclusions can be drawn from our model fits.

\section{Summary and Conclusions}
\Seclabel{Conclusions}

In this paper, we have presented a new method to estimate weak lensing magnification which can be applied to an arbitrary region of space, as expressed in \Eqref{mu_hat_sum_form} and in a more efficient form for computations in \Eqref{mu_hat_efficient_form}. We've shown that the estimator resulting from this method has equivalent statistical power to the optimally-weighted correlation function estimator which has been used for past analyses, and we provide conversions between them in \Eqref{mu_hat_from_w} and \Eqref{w_from_mu_hat}. This conversion also allows a simple way to estimate the errors on measurements made with the correlation function estimator without the use of bootstrap or jackknife methods, which we present in \Eqref{w_std_err}. However, this estimate is likely to be biased low due to the clustering of source galaxies. Our method can calculate a more accurate estimate of error with only a few extra calculations, which presents a large benefit over the requirement of correlation-function-based methods to use bootstrap or jackknife approaches to accurately calculate error. Our method is also capable of handling overlapping lens and source samples, which can increase the signal-to-noise of the magnification signal when this is the case.

The estimator we have presented here provides a suitable mathematical basis to use magnification for mass mapping, for instance through the form in \Eqref{mu_hat_mass_map_form}. However, applying this will be difficult in practice, as the contributions to the magnification signal from actual fluctuations in the projected mass density field are degenerate with localized observational biases, such as due to variable seeing across the field of view. It remains for future research to determine whether this issue can be resolved, through advances in photometric redshift estimation to correct for seeing-induced biases or other techniques.

We've shown a proof-of-principle of our method applied to galaxy-galaxy lensing in the CFHTLenS fields, and shown that the convergence profiles determined with it are consistent with the convergent predictions from mass profiles which are fit to shear data in the fields, as can be seen in \Figref{kappa_with_models}. As expected, in most cases magnification data provides poorer signal-to-noise than does shear, but its contribution is not always negligible. In fact, in certain cases its combination with shear seems to provide significantly smaller errors on model parameters than either alone, likely due to the magnification data helping break degeneracies between multiple parameters. This finding is tentative, however, as our present algorithm was only designed as a proof-of-principle and does not yet include provisions for effects such as dust obscuration, which can result in biases in the magnification signal.

\section*{Acknowledgments}

BRG would like to thank Hendrik Hildebrandt for useful discussions on the methodology presented here and on the details of the photometric redshifts in the CFHTLenS catalogues, Chris Duncan for useful discussions on the methodology, and Thomas Erben for providing clarification on the CFHTLenS field size measurements.

The authors wish to thank the CFHTLenS team for making their galaxy catalogues publicly-available, and Ludo van Waerbeke for supplying us with the convergence maps generated for the CFHTLenS fields.

The authors also wish to thank the anonymous referee for their extensive comments and recommendations on how to improve the paper.

This work is based on observations obtained with MegaPrime/MegaCam, a joint project of CFHT and CEA/DAPNIA, at the Canada-France-Hawaii Telescope (CFHT) which is operated by the National Research Council (NRC) of Canada, the Institut National des Sciences de l'Univers of the Centre National de la Recherche Scientifique (CNRS) of France, and the University of Hawaii. This research used the facilities of the Canadian Astronomy Data Centre operated by the National Research Council of Canada with the support of the Canadian Space Agency. CFHTLenS data processing was made possible thanks to significant computing support from the NSERC Research Tools and Instruments grant program. 

This research uses data from the VIMOS VLT Deep Survey, obtained from the VVDS database operated by Cesam, Laboratoire d'Astrophysique de Marseille, France. 

Funding for the DEEP2 Galaxy Redshift Survey has been provided by NSF grants AST-95-09298, AST-0071048, AST-0507428, and AST-0507483 as well as NASA LTSA grant NNG04GC89G. 

\bibliographystyle{mn2e}
\setlength{\bibhang}{2.0em}
\setlength\labelwidth{0.0em}
\bibliography{full_bib}

\appendix

\section{Derivations of Selected Equations}
\Applabel{derivations}

In \Eqref{mu_stddev}, we presented the expected standard deviation of the estimator $\mu_i$ for the magnification of an individual magnitude bin. This equation can be derived as follows, starting from \Eqref{mu_i_final}:
\begin{align*}
\Eqlabel{mu_stddev_derivation}
\sigma_{\mu_i} &= \frac{\sigma_{dn_i}}{\left|\alpha_i-1\right|dn_{{\rm 0},i}} \\ \numberthis
&= \frac{\sqrt{dn_{i}}}{\left|\alpha_i-1\right|dn_{{\rm 0},i}} \\
&= \frac{\sqrt{dn_{{\rm 0},i}\mu^{\alpha-1}}}{\left|\alpha_i-1\right|dn_{{\rm 0},i}} \\
&\approx \frac{\sqrt{1+(\alpha_i-1)(\mu-1)}}{\left|\alpha_i-1\right|\sqrt{dn_{{\rm 0},i}}}\mathrm{.}
\end{align*}

In \Eqref{w_std_err}, we presented an equation for estimating the standard error of the optimally-weighted correlation function estimator for magnification, without relying on a bootstrap or jackknife approach. This equation can be derived as follows, starting from \Eqref{w_from_mu_hat}:
\begin{align*}
	\Eqlabel{w_std_err_derivation}
	\stderr{w\spr{opt}} = &\;\expec{\left[\alpha-1\right]^2}\stderr{\hat{\mu}} \\ \numberthis
	= &\;\frac{\int_a^b n\sbr{0}(m)\left[\alpha(m)-1\right]^2 dm}{\int_a^b n\sbr{0}(m)dm} \\
	  &\;\times \left(\int_a^b n\sbr{0}(m)\left[\alpha(m)-1\right]^2 dm\right)^{-1/2}  \\
	= &\;\frac{\left(\int_a^b n\sbr{0}(m)\left[\alpha(m)-1\right]^2 dm\right)^{1/2}}{\int_a^b n\sbr{0}(m)dm}  \\
	= &\;\sqrt{ \frac{\expec{\left[\alpha-1\right]^2}}{N\sbr{0}A_{\rm samp}} }\mathrm{,}
\end{align*}
where $N\sbr{0}$ is defined as in \Secref{method_comparison}.

\section{Supplemental Figures}
\Applabel{supplemental_figures}

\begin{figure*}
	\includegraphics[scale=0.9]{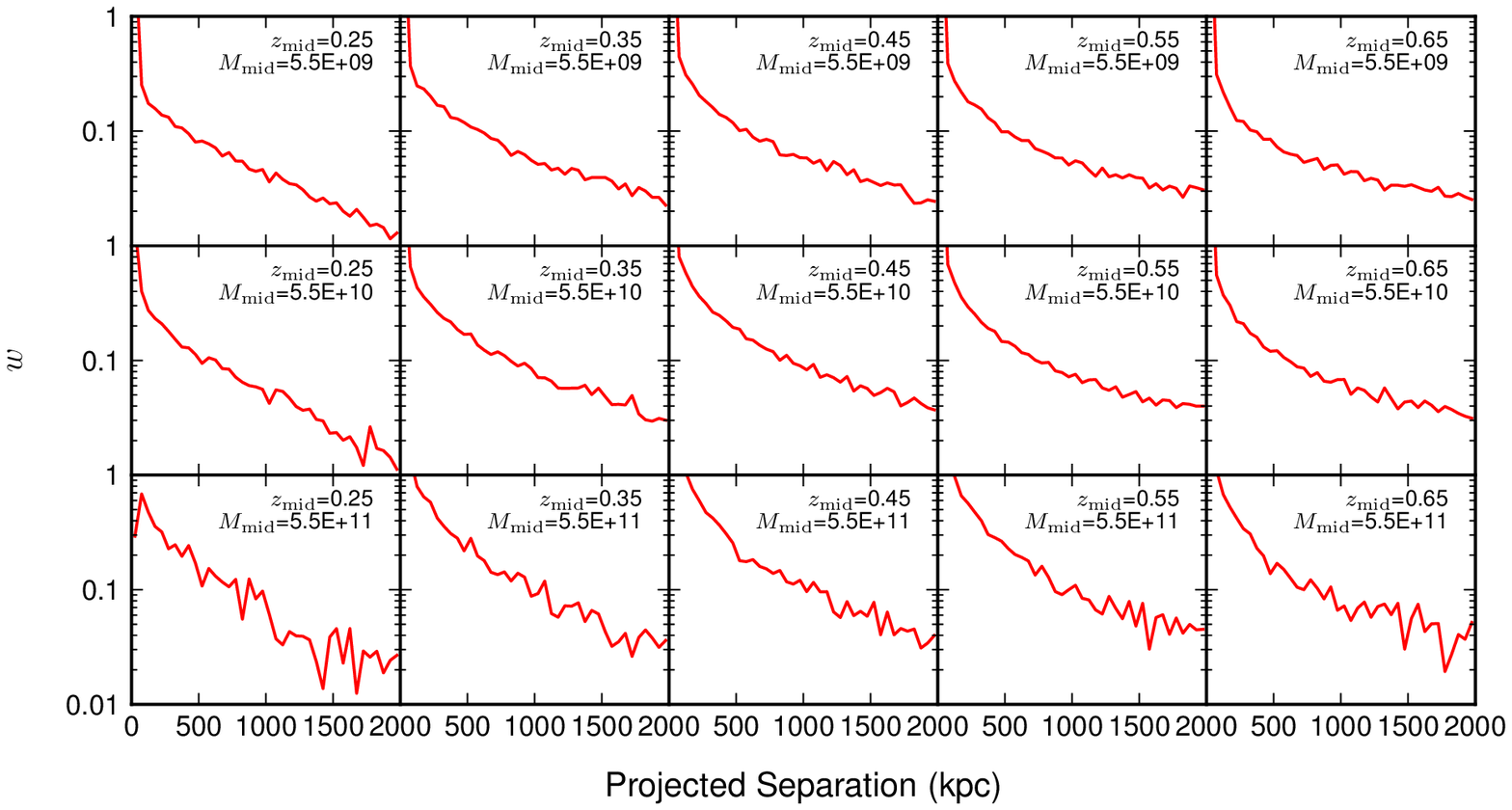}
	\includegraphics[scale=0.9]{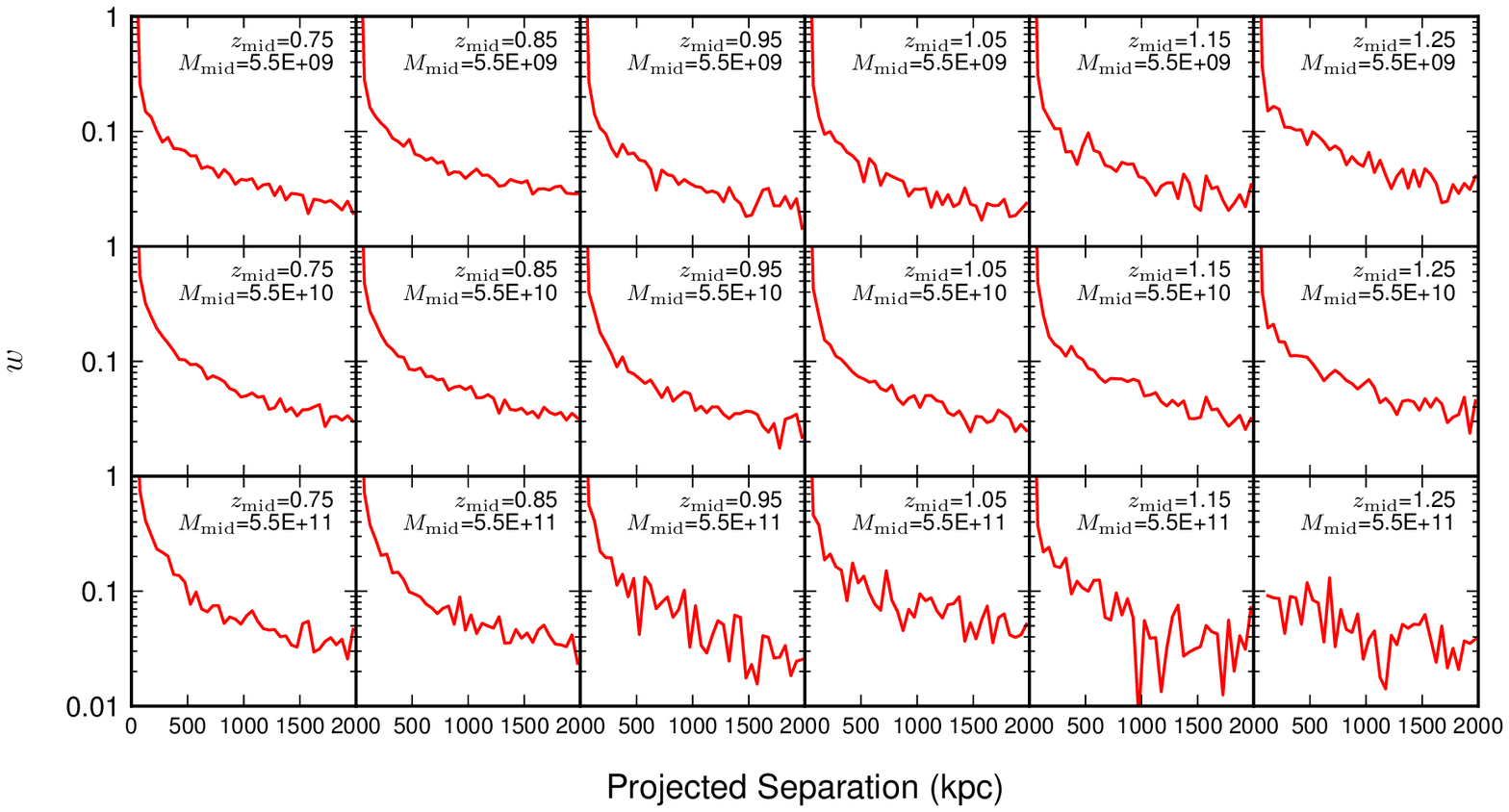}
	\caption[Correlation functions at various lens redshift and mass bins]{Correlation function between galaxies from the lens and source samples in various redshift and stellar mass bins. Low redshift bins are shown in the top panel, and high redshift bins in the bottom panel.}
	\Figlabel{lens_source_same_redshift_corr_funcs}
\end{figure*}

\Figref{lens_source_same_redshift_corr_funcs} shows the correlation function between galaxies in the lens and source samples within the same redshift slices for low-redshift bins. This can be used to estimate the contamination fraction of lenses in the source sample, which we showed in \Figref{lens_contamination_fraction}.

\section{Code Access}
\Applabel{code_access}

In order to facilitate further work with magnification, we make the magnification and shear measurement code used for this paper publicly-available. The code may be accessed online through a repository hosted at \texttt{https://bitbucket.org/brgillis/magnification\_public}. If this repository is no longer available and a web search fails to locate a new host, interested parties are invited to contact the lead author, Bryan Gillis, at \texttt{brg@roe.ac.uk} to request the code.

\label{lastpage}

\end{document}